\documentclass[
 reprint,
 superscriptaddress,   
 altaffilletter,
 amsmath,amssymb,
 aps,
 prd,
 floatfix,
]{revtex4-2}

\usepackage{graphicx}
\usepackage{dcolumn}
\usepackage{bm}
\usepackage{hyperref}
\usepackage[mathlines]{lineno}
\usepackage{mathtools}
\usepackage{comment}
\usepackage{booktabs}
\usepackage{multirow}
\usepackage{longtable}
\usepackage{upgreek}

\newcommand{\rf}[1]{{\color{red}\textbf{$\lceil$REF$\rfloor$}}}

\newcommand{\nubb}{$0\upnu\upbeta\upbeta$}

\newcommand{\xe}[1]{$^{#1}$Xe}

\newcommand{\percent}{\%}

\newcommand{\qbb}{Q$_{\upbeta\upbeta}$}

\newcommand{\ndbd}{$\mathrm{0\upnu\upbeta\upbeta}$}
\newcommand{\code}[1]{\texttt{#1}}
\usepackage{siunitx}
\DeclareSIUnit[number-unit-product = {}]{\inch}{"}
\DeclareSIUnit[number-unit-product = {}]{\angstrom}{Å}
\DeclareSIUnit[number-unit-product = {}]{\bar}{\:bar}
\DeclareSIUnit[number-unit-product = {}]{\mbar}{\:mbar}
\DeclareSIUnit[number-unit-product = {}]{\year}{\:yr}
\DeclareSIUnit\litre{l}
\DeclareSIUnit\liter{l}

\let\onecolumn\onecolumngrid
\renewcommand{\selectlanguage}[1]{}

\begin{document}

\title{Estimating cosmogenic backgrounds in nEXO at SNOLAB using Geant4 and FLUKA}

\date{\today}

\newcommand{\UCSD}{\affiliation{Physics Department, University of California San Diego, La Jolla, CA 92093, USA}}
\newcommand{\McGill}{\affiliation{Physics Department, McGill University, Montr\'eal, QC H3A 2T8, Canada}}
\newcommand{\Stanford}{\affiliation{Physics Department, Stanford University, Stanford, CA 94305, USA}}
\newcommand{\Erlangen}{\affiliation{Erlangen Centre for Astroparticle Physics (ECAP), Friedrich-Alexander University Erlangen-N{\"u}rnberg, Erlangen 91058, Germany}}
\newcommand{\PNNL}{\affiliation{Pacific Northwest National Laboratory, Richland, WA 99352, USA}}
\newcommand{\Carleton}{\affiliation{Department of Physics, Carleton University, Ottawa, ON K1S 5B6, Canada}}
\newcommand{\UMass}{\affiliation{Amherst Center for Fundamental Interactions and Physics Department, University of Massachusetts, Amherst, MA 01003, USA}}
\newcommand{\ITEP}{\affiliation{National Research Center ``Kurchatov Institute'', Moscow, 123182, Russia}}
\newcommand{\LLNL}{\affiliation{Lawrence Livermore National Laboratory, Livermore, CA 94550, USA}}
\newcommand{\UK}{\affiliation{Department of Physics and Astronomy, University of Kentucky, Lexington, KY 40506, USA}}
\newcommand{\BNL}{\affiliation{Brookhaven National Laboratory, Upton, NY 11973, USA}}
\newcommand{\SLAC}{\affiliation{SLAC National Accelerator Laboratory, Menlo Park, CA 94025, USA}}
\newcommand{\RPI}{\affiliation{Department of Physics, Applied Physics, and Astronomy, Rensselaer Polytechnic Institute, Troy, NY 12180, USA}}
\newcommand{\Laurentian}{\affiliation{School of Natural Sciences, Laurentian University, Sudbury, ON P3E 2C6, Canada}}
\newcommand{\IHEP}{\affiliation{Institute of High Energy Physics, Chinese Academy of Sciences, Beijing, 100049, China}}
\newcommand{\IME}{\affiliation{Institute of Microelectronics, Chinese Academy of Sciences, Beijing, 100029, China}}
\newcommand{\Sherbrooke}{\affiliation{Universit\'e de Sherbrooke, Sherbrooke, QC J1K 2R1, Canada}}
\newcommand{\Alabama}{\affiliation{Department of Physics and Astronomy, University of Alabama, Tuscaloosa, AL 35405, USA}}
\newcommand{\UNCW}{\affiliation{Department of Physics and Physical Oceanography, University of North Carolina Wilmington, Wilmington, NC 28403, USA}}
\newcommand{\UBC}{\affiliation{Department of Physics and Astronomy, University of British Columbia, Vancouver, BC V6T 1Z1, Canada}}
\newcommand{\TRIUMF}{\affiliation{TRIUMF, Vancouver, BC V6T 2A3, Canada}}
\newcommand{\Drexel}{\affiliation{Department of Physics, Drexel University, Philadelphia, PA 19104, USA}}
\newcommand{\ORNL}{\affiliation{Oak Ridge National Laboratory, Oak Ridge, TN 37831, USA}}
\newcommand{\CSU}{\affiliation{Physics Department, Colorado State University, Fort Collins, CO 80523, USA}}
\newcommand{\Yale}{\affiliation{Wright Laboratory, Department of Physics, Yale University, New Haven, CT 06511, USA}}
\newcommand{\USD}{\affiliation{Department of Physics, University of South Dakota, Vermillion, SD 57069, USA}}
\newcommand{\Mines}{\affiliation{Department of Physics, Colorado School of Mines, Golden, CO 80401, USA}}
\newcommand{\CUP}{\affiliation{IBS Center for Underground Physics, Daejeon, 34126, South Korea}}
\newcommand{\UWC}{\affiliation{Department of Physics and Astronomy, University of the Western Cape, P/B X17 Bellville 7535, South Africa}}
\newcommand{\SUBATECH}{\affiliation{SUBATECH, Nantes Universit\'e, IMT Atlantique, CNRS/IN2P3, Nantes 44307, France}}
\newcommand{\Caltech}{\affiliation{California Institute of Technology, Pasadena, CA 91125, USA}}
\newcommand{\Bern}{\affiliation{LHEP, Albert Einstein Center, University of Bern, 3012 Bern, Switzerland}}
\newcommand{\Skyline}{\affiliation{Skyline College, San Bruno, CA 94066, USA}}
\newcommand{\SNOLAB}{\affiliation{SNOLAB, Lively, ON P3Y 1N2, Canada}}
\newcommand{\Muenster}{\affiliation{Institut f{\"u}r Kernphysik, Westf{\"a}lische Wilhelms-Universit{\"a}t M{\"u}nster, M{\"u}nster 48149, Germany}}
\newcommand{\FRIB}{\affiliation{Facility for Rare Isotope Beams, Michigan State University, East Lansing, MI 48824, USA}}
\newcommand{\Queens}{\affiliation{Department of Physics, Queen's University, Kingston, ON K7L 3N6, Canada}}
\newcommand{\Windsor}{\affiliation{Department of Physics, University of Windsor, Windsor, ON N9B 3P4, Canada}}
\newcommand{\TUM}{\affiliation{Physikdepartment and Excellence Cluster Universe, Technische Universit{\"a}t M{\"u}nchen, Garching 80805, Germany}}
\newcommand{\Montclaire}{\affiliation{Department of Physics and Astronomy, Montclair State University, Montclair, NJ 07043, USA}}
\newcommand{\McMaster}{\affiliation{Department of Physics and Astronomy, McMaster University, Hamilton, ON L8S 4M1, Canada}}
\newcommand{\SJTU}{\affiliation{School of Physics and Astronomy, Shanghai Jiao Tong University, Shanghai 200240, China}}
\newcommand{\Hawaii}{\affiliation{Department of Physics and Astronomy, University of Hawaii at Manoa, Honolulu, HI 96822, USA}}
\newcommand{\KEK}{\affiliation{KEK, High Energy Accelerator Research Organization 1-1 Oho, Tsukuba, Ibaraki 305-0801, Japan}}
\newcommand{\SFU}{\affiliation{Simon Fraser University, Burnaby, BC, V5A 1S6, Canada}}


\author{R.~Ross}\McGill
\author{S.~Al Kharusi}\altaffiliation{Now at: Physics Department, Stanford University, Stanford, CA 94305, USA}\McGill
\author{T.~Brunner}\McGill\TRIUMF
\author{S.~Sangiorgio}\LLNL

\author{E.~Caden}\SNOLAB\McGill\Laurentian
\author{U.~Wichoski}\Laurentian\Carleton

\author{B.~G.~Lenardo}\SLAC
\author{B.~Mong}\SLAC
\author{A.~Odian}\SLAC

\author{A.~Piepke}\Alabama

% ...

\begin{abstract}
The nEXO experiment was designed to search for neutrinoless double beta decay in the isotope $^{136}$Xe, using a five-tonne, isotopically-enriched liquid xenon time projection chamber (TPC).
To meet nEXO's sensitivity goal, significant effort is placed on background mitigation strategies.
Here, we evaluate the expected cosmogenic background rate for nEXO if placed at its preferred underground host site, SNOLAB.
A comparison of results from Geant4 and FLUKA Monte Carlo simulations are presented. We perform detailed Monte Carlo simulations of muons reaching the underground SNOLAB environment using both Geant4 and FLUKA to study the production of radioactive isotopes from both direct spallation reactions or subsequent neutron-induced reactions. We find that, of the isotopes produced, the only one of concern for nEXO is \xe{137}, produced by neutron capture on the xenon target. We develop a veto strategy that achieves an anticipated 92\% \xe{137} rejection efficiency while keeping dead time to less than 1 day over a 10 year exposure.
\end{abstract}

\maketitle

\section{Introduction}
\label{sec:intro}


Neutrinoless double beta decay (\nubb{}) is a hypothesized nuclear decay that would violate the conservation of lepton number, and provide evidence for Majorana neutrino masses \cite{majorana1937teoria,schechter1982neutrinoless}, providing new pathways for mass generation of neutrinos in extensions to the Standard Model~\cite{dolinski2019neutrinoless}. The proposed nEXO experiment is a search for \nubb{} in the isotope xenon-136, targeting a half-life exclusion sensitivity of \SI{1.35e28}{\year} at \SI{90}{\percent} C.L.\ after \SI{10}{years}, probing much of the neutrino mass parameter space associated with inverted ordering \cite{adhikari2021nexo}.

nEXO is designed to use \SI{5000}{\kilo\gram} of liquid xenon (LXe), enriched to \SI{90}{\percent} \xe{136} with \SI{3648}{kg} within the drift region of a monolithic cylindrical time projection chamber (TPC). Scintillation light and ionization charge provide measurements of event energy, three-dimensional position, and topology, ie. an event may deposit one or more resolvable clusters of energy~\cite{nEXO_SiPM_testing,li2019simulation}. These observables distinguish the localized, uniformly distributed \nubb{} signal at \qbb{}$=2457.83(37)$~keV \cite{RedshawMassDoubleBetaDecay2007} from backgrounds. The baseline energy resolution is \SI{1}{\percent} FWHM at \qbb{}, corresponding to the energy window (2445.5, 2470.1) keV, here referred to as the energy region-of-interest (EROI). 

The TPC vessel and field cage are designed primarily with ultra-pure electroformed copper~\cite{HOPPE2014116}, with the instrumentation and additional internal material making up a small fraction of the total mass. The TPC will be surrounded by a spherical, vacuum-insulated cryostat filled with a cryogenic heat-transfer fluid, HFE-7200, which also serves as a radiation shield. In this work, we assume the cryostat vessel is made of nickel. Moving outward, a water-filled outer detector (OD) will add passive shielding, and be instrumented as an active muon veto.  A detailed accounting of nEXO's design and material budget is given in Ref~\cite{adhikari2021nexo}.

The preferred host site for nEXO is SNOLAB, where the \SI{2}{\kilo\meter} overburden of rock reduces the flux of cosmic-ray muons to approximately \num{e-7} times that at sea-level, resulting in a total integrated muon flux of \\  \SI{0.29}{\per\square\meter\per\day} as measured by SNO~\cite{SNOLAB_muon_flux}. Residual cosmic muon flux can create so-called \textit{cosmogenic} backgrounds. 

This work evaluates nEXO's cosmogenic backgrounds; specifically, those induced by cosmic muons and generated \textit{in situ} at SNOLAB during data taking. Activation from above-ground material exposure is outside its scope. In the next section, we discuss what constitutes a cosmogenic background; in Section \ref{sec:config_and_sim} we present the simulated nEXO configurations and virtual muon source; Section \ref{sec:activation_rates} describes the method for estimating activation rates and uncertainties using two independent simulations; Section \ref{sec:ROCs} describes how background radionuclides are filtered from total yields; and Section \ref{sec:coincidence_veto_xe137} discusses how the dominant, \xe{137} background might be mitigated.

\section{Cosmogenic Backgrounds}
\label{sec:cosmogenic_backgrounds_intro}
In the upper atmosphere, pion and kaon decays produce muons of sufficiently high energies to penetrate significant depths of the Earth's crust \cite{GaisserCosmicRays}. Underground, muon-induced fission and spallation interactions continually produce radionuclides and free neutrons (which can be captured, producing other radioisotopes) \cite{Gratta_FLUKA_Neutrons}. Cosmogenic backgrounds are those due to muon-induced interactions with local material leading directly or indirectly to energy deposits in detector media. 

Here we refer to nuclides produced via mechanisms other than muon induced neutron capture as spallation products; for instance, those arising from nuclear fragmentation, isotopes of I and Te produced by removing nucleons from Xe, or those from charge-exchange interactions like $^{136}$Xe$(p,n)$$^{136}$Cs. 

Backgrounds of concern to nEXO are those which contribute events with energies in the EROI that appear as single-sited (SS) deposits occurring throughout the LXe thereby being indistinguishable in their event topology from \nubb{}. These backgrounds are uncorrelated in time with the passage of muons. Cosmogenically activated isotopes that decay through prompt $(\mathrm{n},\upalpha)$ or $(\mathrm{n},\upgamma)$ do so on sufficiently short timescales that a TPC-wide veto of $1 \mathrm{s}$ following each muon passage is sufficient to reject their contributions to the \nubb{} signal region with a negligible loss to exposure.

Generally, $\upbeta$-decaying isotopes have longer lifetimes. Vetoing the full TPC for tens of minutes each time a muon is detected would cost considerable exposure time resulting in a significant net loss to nEXO's sensitivity to \nubb{}. Thus, a more selective veto scheme is needed to minimize deadtime and maximize exposure to possible \nubb{} candidate events.

This study follows a similar approach to the EXO-200 experiment in identifying cosmogenic backgrounds \cite{Albert2016-cosmogenic}: we simulate a muon flux impinging upon nEXO and its surrounding material at SNOLAB, tallying up the nuclear species created. The methods used to simulate SNOLAB's cosmic muon flux and accurately represent nEXO in simulation are in Section \ref{sec:config_and_sim}. Of all isotopes produced through cosmogenic activation, only a subset are ``radionuclides of concern'' (ROCs). The selection of ROCs is outlined in Section \ref{sec:ROCs}.

\section{Geometry and Muon Source}
\label{sec:config_and_sim}
\subsection{Geometry}

\begin{figure*}[h]
    \centering
    \begin{minipage}[c]{0.68\textwidth}
        \centering
        \includegraphics[width=\linewidth]{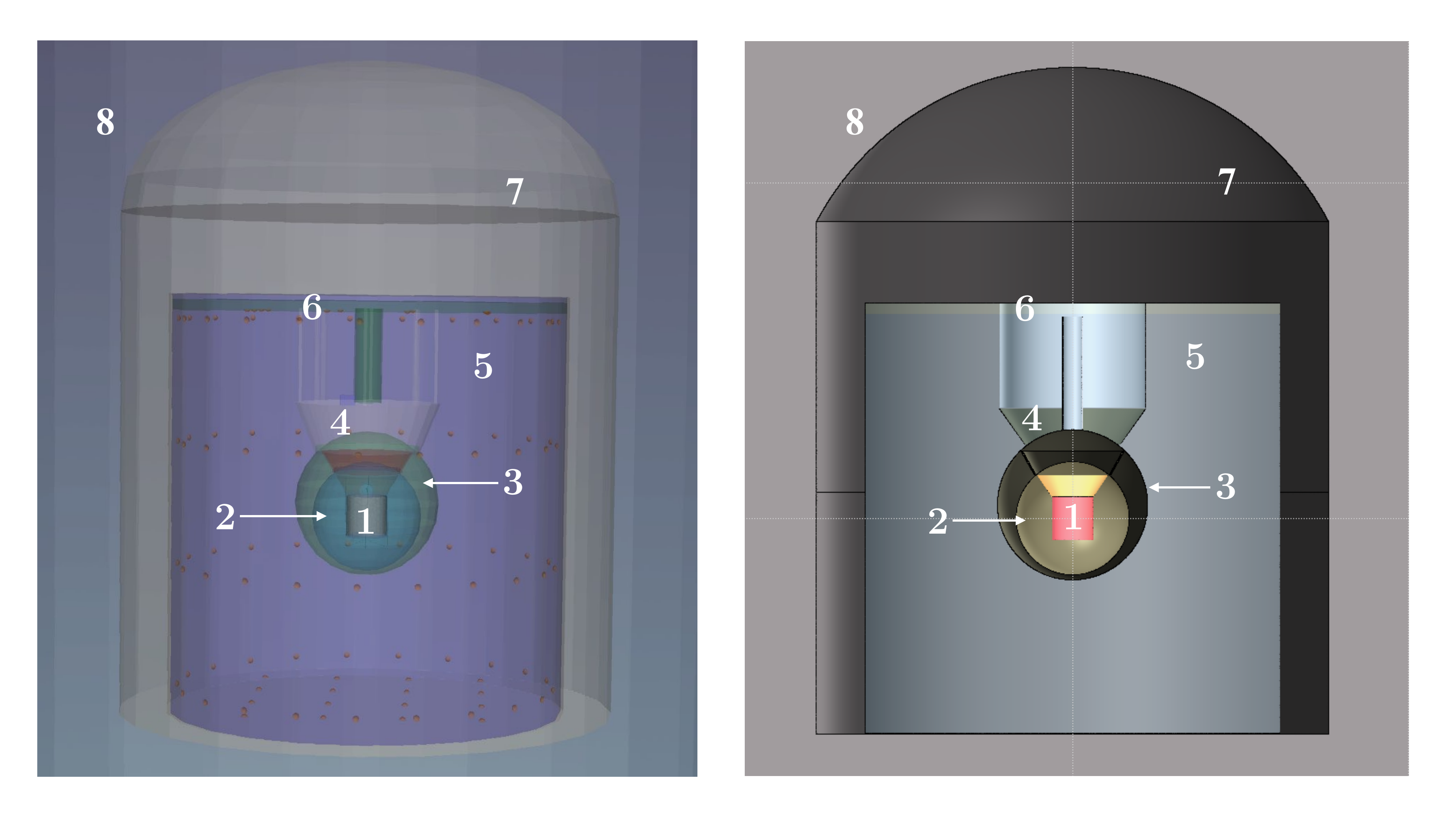}
    \end{minipage}
    \begin{minipage}[c]{0.3\textwidth}
        \centering
        \begin{tabular}{cl}
            \hline
            \textbf{No.} & \textbf{Volume} \\
            \hline
            1 & LXe-filled copper TPC \\
            2 & HFE-filled inner cryostat \\
            3 & Outer cryostat \\
            4 & Copper supports/feedthroughs \\
            5 & Ultra-pure water-filled OD \\
            6 & Nitrogen cover gas \\
            7 & SNOLAB Cryopit \\
            8 & Norite rock \\
            \hline
        \end{tabular}
    \end{minipage}

    \caption{nEXO configuration as implemented in \code{Geant4} (left) and \code{FLUKA} (right) with major volumes indicated.}
    \label{fig:nEXO_Config}
\end{figure*}

The geometries implemented in \code{FLUKA} and \code{Geant4} contain the important features of nEXO while neglecting the minutiae. A visualization of the configurations deployed side-by-side is shown in Figure \ref{fig:nEXO_Config}. The overall material masses in \code{FLUKA} are within \SI{0.1}{\percent} of those in the \code{Geant4} geometry. Component dimensions in the \code{FLUKA} simulation match their \code{Geant4} counterparts as closely as possible, but small components are somewhere included as mass additions to larger components, or neglected altogether, as is the case for the PMTs lining the OD. 

\begin{table*}[h]
\centering
\begin{tabular}{
    l
    l
    l
    S
    S
    S
}
\toprule
Category & Name & Material & {FLUKA Mass (kg)} & {GEANT4 Mass (kg)} & {\% Difference} \\
\midrule

\multirow{3}{*}{TPC}
& Xenon & LXe & 4.963e3 & 4.963e3 & 0.005 \\
& Vessel & Copper & 7.236e2 & 7.235e2 & -0.024 \\
& Support Cone & Copper & 8.767e1 & 8.771e1 & 0.050 \\

\midrule

\multirow{3}{*}{Inner Cryostat}
& HFE & HFE-7200 & 3.147e4 & 3.148e4 & 0.045 \\
& Shell & Nickel & 2.555e3 & 2.557e3 & 0.063 \\
& Support Cone & Nickel & 2.128e3 & 2.128e3 & 0.013 \\

\midrule

\multirow{4}{*}{Outer Cryostat}
& Shell & Nickel & 8.011e3 & 8.010e3 & -0.004 \\
& Support & Nickel & 2.993e3 & 2.994e3 & 0.033 \\
& OD Support & Stainless Steel & 1.850e3 & 1.850e3 & 0.033 \\
& Feedthrough & Stainless Steel & 2.156e3 & 2.155e3 & -0.055 \\

\midrule

\multirow{2}{*}{Outer Detector}
& Water & Water & 1.483e6 & 1.483e6 & 0.001 \\
& Tank & Stainless Steel & 7.154e4 & 7.155e4 & 0.019 \\

\bottomrule
\end{tabular}

\caption{Comparison of significant simulated masses.}
\label{tab:mass_comparison}

\end{table*}

\subsection{Physics Models}

In the \code{Geant4} simulations we use the \code{Shielding} physics list, which includes models for neutron transport down to thermal energies. Details on the included processes and interaction models for \code{Shielding} can be found in \cite{Geant4PhysicsReference}, including the energy ranges of each model. The cross-sections for neutron interactions are found in the \code{G4NDL 4.6} library \cite{mendoza2018update} that is based on the \code{JEFF-3.3} nuclear data files \cite{plompen2020joint}. 

In \code{FLUKA}, the most relevant physics options enabled are outlined in Table \ref{tab:fluka_physics}. These include so-called ``precision'' defaults which activate muon photonuclear events and neutron transport down to thermal energies. Some other options enabled were consequently redundant but are listed for clarity. Cross-sections for low-energy neutron interactions were taken from the \code{JEFF-3.3} database \cite{plompen2020joint} as linked through \code{LOW-PWXS} options. Due to the high energies of the simulated muons, the \code{DPMJET} event generator \cite{DPMJET-III,DPMJET-III-2} was explicitly requested at compile time for high-energy interactions ($>$ \SI{5}{\giga\electronvolt\per\text{nucleon}}). The \code{RQMD} event generator was also used for  $\mathrm{E} \in (0.1,5)$ \SI{}{\giga\electronvolt\per\text{nucleon}} \cite{rQMD_1,rQMD_2}.

\subsection{Simulating Muons at SNOLAB}
\subsubsection*{Energy and Angular Distributions}



Simulations were performed with dedicated \\ \code{Geant4 v10.7} \cite{agostinelli_geant4simulation_2003} and \code{FLUKA v4-5.0} \cite{FLUKA,FLUKAlatest}  Monte Carlo (MC) implementations that were run with nEXO in the SNOLAB Cryopit. The MC sampling procedures for both \code{FLUKA} and \code{Geant4} implementations were identical and account for the correlated energy and angular distribution of muons. 

We follow a muon simulation scheme similar to \cite{CollaborationCosmogenicNeutron2025}. The muon energy distribution at SNOLAB is assumed to follow that described by Lipari and Stanev in \cite{Lipari_Stanev_1991} which is a function of the zenith angle via the slant-depth, $h = h_0/\cos\theta$, where $h_0$ is the approximately \SI{6}{\kilo\metre} water equivalent vertical depth of SNOLAB:
\begin{equation}
    \centering
    \label{eq:gaisser_muon_energies}
    \frac{dN}{dE_\mu} = Ae^{-bh(\gamma_\mu - 1)}\cdot\left[E_\mu + \epsilon_\mu(1 - e^{-bh})\right]^{-\gamma_\mu} ,
\end{equation}
where $A$ is a normalization constant. Both $\epsilon$ and $b$ are medium-dependent with $\epsilon$ being the critical muon energy, and $b$ the attenuation coefficient with contributions from bremsstrahlung, pair production, and photonuclear events \cite{Groom_Mokhov_Striganov_2001}. $\gamma_\mu$ is the spectral index. 

Muon energies were sampled up to \SI{25}{\tera\electronvolt}, producing a mean energy of \SI{363}{\giga\electronvolt}. This is consistent with measurements at SNOLAB~\cite{aharmim2019cosmogenic}. The muon charge ratio was fixed to $\mu_+/\mu_- = 1.3$ \cite{darwin_cosmogenics}. The underground muon zenith angles were sampled from the Mei and Hime parameterization with random, uniformly distributed azimuthal angles consistent with the flat overburden at SNOLAB:

\begin{equation}
    \centering
    \label{eq:mei_hime_zenith_distribution}
    I(h_0, \theta) = \frac{\left[ I_1e^{(-h_0\sec\theta/\lambda_1)} + I_2e^{(-h_0\sec\theta/\lambda_2)}\right]}{\cos\theta} ,
\end{equation}

where $I_1$, $I_2$, $\lambda_1$, and $\lambda_2$ are fit parameters of equation 3 given in \cite{MeiHime}. 

\subsubsection*{Muon Generation and Targeting Scheme}
\label{sec:mu_generation}

\begin{figure*}[]
    \centering
    \includegraphics[width=.7\linewidth]{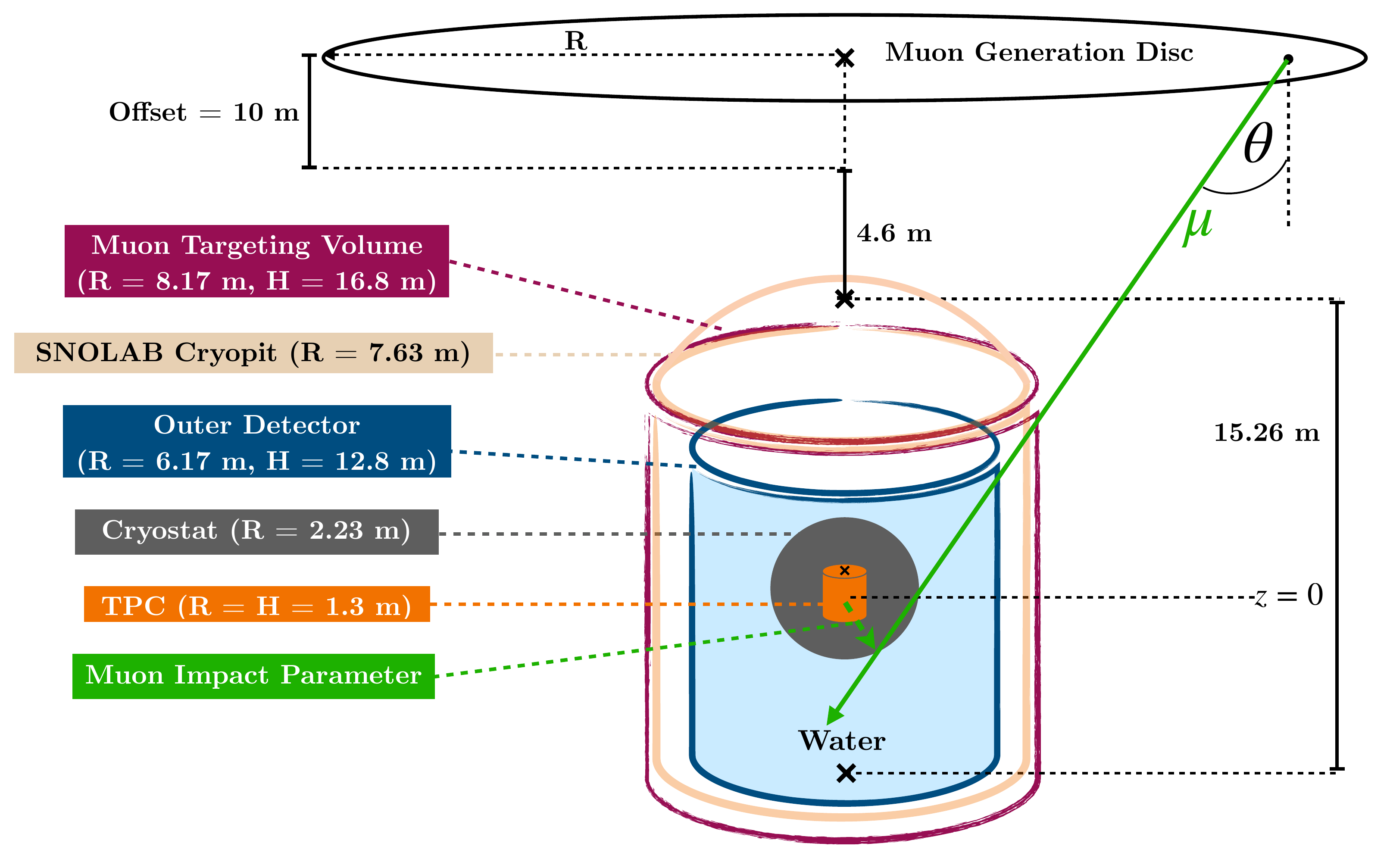}
    \caption{Illustration of parameters used for the generation of muons in Geant4 and FLUKA including the virtual muon target volume (maroon), muon generation disk, and basic concentric elements of the nEXO configuration. Not drawn to scale.}
    \label{fig:muon-generation}
    \end{figure*}

The procedure for simulating the muon flux at SNOLAB followed similar work done for EXO-200 \cite{Albert2016-cosmogenic}. A large disk, concentric with the nEXO TPC was placed above the cryopit and inlaid into the norite rock ($\rho = $ \SI{2.894}{\gram\per\cubic\centi\metre}~\cite{Scallon_2019}) such that the minimum distance muons traveled through the rock was \SI{10}{\meter}. Muon initial positions were selected on the circular disk uniformly. Their azimuthal angles were randomly sampled, and their zenith angles sampled from
\begin{equation}
    \label{eqn:mhpdf}
    \frac{dI(h_0, \theta)}{d\Omega} = AI(h_0, \theta)\sin\theta,
\end{equation}
where $A$ is a normalization constant.

Of all the muons generated with the above procedure, only those intersecting a target volume were propagated in the simulations. The muon target volume is a cylinder concentric with the nEXO OD, but \SI{2}{\meter} larger in each direction (radius + \SI{2}{\meter}, height + \SI{4}{\meter}). This is illustrated in Figure~\ref{fig:muon-generation}. Assuming straight line trajectories for the muons, their intersection points with the target surface were calculated by solving a set of parametric equations. Muons missing the target volume were rejected and their direction vectors, initial positions, and energies were resampled. 

The likelihood of a muon to create  radionuclides of concern (ROCs) or other secondaries of concern decreases markedly with its track's distance from the TPC. The norite rock at SNOLAB has a nuclear interaction length $\lambda_n \approx$ \SI{50}{\centi\meter} and a Moliere radius of $X_0 \approx$ \SI{10}{\centi\meter}~\cite{Groom_Mokhov_Striganov_2001}. Muon-induced hadronic showers separated from the detector by meters of rock are therefore unlikely to contribute to backgrounds. Neutrons produced in hadronic showers meters away would have to scatter into the solid angle subtended by the TPC \textit{and} make it through the rock, water, and cryostat before interacting.

Simulating only muons that pass within a larger concentric target volume is therefore sufficient provided the correlations between zenith angle and energy are preserved, and no other correlations are inadvertently created through the targeting selection. The targeting scheme is illustrated in Figure \ref{fig:muon-generation}. Comparable simulation parameters for both \code{FLUKA} and \code{Geant4} are listed in Table \ref{tab:mass_comparison}.

The muon data phase space files were read in by the custom \code{FLUKA} source routine. For both \code{Geant4} and \code{FLUKA}, we generate datasets of approximately \num{2e7} muons impinging upon the target volume. This corresponds to an exposure time of over \SI{600}{yr} at SNOLAB.


\section{Activation Rates}
\label{sec:activation_rates}

The activation rates in this work were estimated with independent simulations performed with \code{Geant4} and \code{FLUKA}. We present combined rates for the production of nuclides. These are determined as the simple mean of both the \code{Geant4} and \code{FLUKA} rates when there are nonzero production rates from both simulations:
\begin{align}
\label{eqn:activation_rate}
R_q &= \frac{G(q) + F(q)}{2},
\end{align}
where $G(q) > 0$ and $F(q) > 0$ are the mean activation rates for isotope $q$ as determined by \code{Geant4} and \code{FLUKA}, respectively. In the case that either the \code{Geant4} or \code{FLUKA} activation rate is 0, we conservatively accept the non-zero rate as the combined rate.

We assume uncorrelated fluctuations between \code{Geant4} and \code{FLUKA}, and accordingly estimate the statistical uncertainty by propagating their independent Monte Carlo uncertainties:
\begin{align}
\sigma_{\mathrm{stat}}(R_q) = \frac{1}{2}\sqrt{\sigma_G^2 + \sigma_F^2},
\end{align}
where $\sigma_G$ and $\sigma_F$ are the statistical uncertainties associated with $G(q)$ and $F(q)$, respectively.

The systematic uncertainty is approximated by the quadrature sum of a model uncertainty, taken as half the difference between the \code{Geant4} and \code{FLUKA} rates, and a contribution from the uncertainty in the muon flux rate at SNOLAB used to determine exposure time:
\begin{equation}
\label{eqn:activation_systematic_uncertainty}
\sigma_{\mathrm{syst}}(R_q) =
\sqrt{
\left(\frac{G(q) - F(q)}{2} \right)^{2}
+ R_q^2 \frac{\sigma^2(\phi_\mu)}{\phi_\mu^2}
},
\end{equation}
where the muon flux through the target volume, $\phi_\mu$, has a standard uncertainty $\sigma(\phi_\mu)$.



\section{Radionuclides of Concern}
\label{sec:ROCs}
Of the many cosmogenic nuclides produced by muon spallation in the simulations, very few are estimated potential background sources to nEXO. Following \cite{adhikari2021nexo}, we define ROCs as the cosmogenic radionuclides that result in more than 0.01 SS-like events in the EROI, within the central \SI{2000}{\kg} of LXe, per year. This shifts our focus onto only a handful from over 300 distinct cosmogenic nuclides produced in simulation. The methods for filtering the ROCs from the rest are outlined in this section.

\subsubsection*{Prompt Muon Veto}

The nEXO design uses a prompt muon veto to reject signals from neutron capture cascades and very-short-lived radioisotopes. The signals from muons passing through the TPC can be easily rejected due to their high energies (\SI{2}{\mega\electronvolt\per\cm} of track). For muons which do not pass through the TPC, we use the instrumented outer shield as a tracking detector to tag the passage of muons and apply a prompt veto, defined as 1 second. We therefore cut the contribution of radionuclides with half-lives shorter than \SI{100}{\milli\s}.

\subsubsection*{Fiducial Volume}

The fiducial volume (FV) is defined as the volume that is $>2$~cm from the inner surfaces of the TPC field cage; this is the volume in which the electric field is anticipated to be highly uniform and contains 3281~kg of LXe. The surrounding LXe, including that contained inside the vessel but outside the TPC field cage, is referred to as the skin and acts as an active shield~\cite{nEXO:2020mec}. The copper radionuclides are bound to the copper TPC vessel and field-shaping rings. A fiducial volume cut can entirely eliminate their $\upbeta$-decay background contribution as electrons with MeV energies are strongly attenuated by the LXe with sub-cm mean ranges \cite{NIST_ESTARElectrons}. Prompt gammas from copper and cobalt are rejected via the prompt muon veto. 

To estimate backgrounds from components outside the LXe, we use hit efficiencies from the \code{Geant4} simulations described in \cite{adhikari2021nexo}, where hit efficiency is defined as the probability for a given decay to produce a signal in the \ndbd{} region of interest \cite{adhikari2021nexo}.

\subsubsection*{Topological Discrimination}
The fraction of delayed gammas resulting from nuclides produced in the copper vessel surviving MS cuts was found to be negligible. Radionuclides decaying via $\beta^+$ emission and electron capture have a meager 1.4\% chance to be tagged as SS events. Accordingly, $\beta^+$ and electron capture -decaying species must be produced in the TPC at rates exceeding \SI{0.5}{\year^{-1}} to constitute a background concern; we found none that meet this threshold.

Coincident $^{60}\mathrm{Co}$ gamma pairs have a negligible background contribution due to cosmogenic activity alone. At a mean activation rate of \SI{0.045}{\year^{-1}} in the TPC vessel copper, the fraction of which whose gammas are collinear, and enter the FV, and make a SS deposit within the EROI, is vanishingly small. We therefore turn our focus inward to the LXe.

\subsubsection*{Resulting ROCs}
Applying these cuts to the list of spallation products in the simulations produces a small list of ROCs whose production rates exceed \SI{0.01}{\year^{-1}}. This list is shown in Table \ref{tab:isotope-rates}. Of these, \xe{137} is the only species found to result in a potential background contribution. The total production rate of all other ROC in the LXe amounts to fewer than 2 counts per year at SNOLAB, a negligible fraction of which are anticipated to survive the multi-dimensional nEXO analysis cuts to produce background signals in the EROI. We also list $^{20}\mathrm{F}$ and $^{16}\mathrm{N}$ as we have found them to have markedly higher production rates than other radioisotopes in the HFE, though still insufficient to produce backgrounds.


\begin{table*}[t]
\centering
\caption{Production rates of ROC in both \code{FLUKA} and \code{Geant4} listed with statistical and systematic uncertainties, respectively. Where a nuclide is present in one simulation, but not the other, we conservatively take the non-zero result. Here we indicate a 0 result with ``—''.}
\label{tab:isotope-rates}
\begin{tabular}{@{}llllll@{}}
\toprule
Nuclide & Half-life & Region & Energy [keV] & \code{FLUKA} Rate [$\mathrm{yr}^{-1}$] & \code{Geant4} Rate [$\mathrm{yr}^{-1}$] \\
\midrule
Xe-137 & 3.82 min & LXe & 4.16e+03 & $10.4 \pm 0.1 \pm 0.4$ & $14.8 \pm 0.8 \pm 0.6$ \\
I-134 & 52.5 min & LXe & 4.08e+03 & $(3.67 \pm 0.23 \pm 0.14) \times 10^{-1}$ & $(3.31 \pm 0.34 \pm 0.12) \times 10^{-1}$ \\
I-135 & 6.58 h & LXe & 2.63e+03 & $(2.86 \pm 0.20 \pm 0.11) \times 10^{-1}$ & $(3.26 \pm 0.33 \pm 0.12) \times 10^{-1}$ \\
I-132 & 2.29 h & LXe & 3.58e+03 & $(3.28 \pm 0.22 \pm 0.12) \times 10^{-1}$ & $(2.71 \pm 0.30 \pm 0.10) \times 10^{-1}$ \\
I-130 & 12.4 h & LXe & 2.94e+03 & $(3.19 \pm 0.21 \pm 0.12) \times 10^{-1}$ & $(2.48 \pm 0.28 \pm 0.09) \times 10^{-1}$ \\
I-136 & 1.39 min & LXe & 6.88e+03 & $(2.58 \pm 0.61 \pm 0.10) \times 10^{-2}$ & $(9.04 \pm 1.60 \pm 0.34) \times 10^{-2}$ \\
Sb-129 & 4.37 h & LXe & 2.38e+03 & --- & $(5.43 \pm 1.20 \pm 0.20) \times 10^{-2}$ \\
Te-131 & 25 min & LXe & 2.23e+03 & $(3.44 \pm 0.70 \pm 0.13) \times 10^{-2}$ & $(7.24 \pm 1.40 \pm 0.27) \times 10^{-2}$ \\
Sb-132 & 2.79 min & LXe & 5.55e+03 & --- & $(3.36 \pm 0.95 \pm 0.13) \times 10^{-2}$ \\
Sb-126 & 12.3 d & LXe & 3.67e+03 & --- & $(3.10 \pm 0.91 \pm 0.12) \times 10^{-2}$ \\
Sb-131 & 23 min & LXe & 3.23e+03 & --- & $(3.10 \pm 0.91 \pm 0.12) \times 10^{-2}$ \\
Sb-124 & 60.2 d & LXe & 2.91e+03 & $(2.58 \pm 0.61 \pm 0.10) \times 10^{-2}$ & $(3.36 \pm 0.95 \pm 0.13) \times 10^{-2}$ \\
Te-133 & 12.5 min & LXe & 2.92e+03 & $(1.86 \pm 0.52 \pm 0.07) \times 10^{-2}$ & $(3.36 \pm 0.95 \pm 0.13) \times 10^{-2}$ \\
Sb-130 & 39.5 min & LXe & 5.07e+03 & --- & $(2.07 \pm 0.74 \pm 0.08) \times 10^{-2}$ \\
Sb-133 & 2.34 min & LXe & 4.01e+03 & --- & $(1.81 \pm 0.69 \pm 0.07) \times 10^{-2}$ \\
Cs-136 & 13.2 d & LXe & 2.55e+03 & $(1.43 \pm 0.45 \pm 0.05) \times 10^{-2}$ & $(2.07 \pm 0.74 \pm 0.08) \times 10^{-2}$ \\
Sb-128 & 9.05 h & LXe & 4.36e+03 & --- & $(1.55 \pm 0.64 \pm 0.06) \times 10^{-2}$ \\
\bottomrule
F-20 & 11.1 s & HFE & 7.02e+03 & $12.2 \pm 0.1 \pm 0.5$ & $30.0 \pm 1.6 \pm 1.1$ \\
N-16 & 7.13 s & HFE & 3.73e+06 & $7.55 \pm 0.10 \pm 0.28$ & $11.1 \pm 0.6 \pm 0.4$ \\
\bottomrule
\end{tabular}
\end{table*}

\section{Coincidence Veto for $^{137}\mathrm{Xe}$}
\label{sec:coincidence_veto_xe137}

Of all the listed ROCs, $^{137}$Xe is the dominant background contributor. It is produced by the neutron capture reaction $^{136}$Xe$(\mathrm{n},\upgamma)^{137}$Xe*, where the excited daughter promptly relaxes through a $\upgamma$ cascade with a summed energy of 4.025 MeV\cite{Albert2016_nCaptureXe136}. \xe{137} has a half-life of \SI{3.818(0.013)}{\minute} \cite{Browne_Tuli_2007}. Convection and flow of LXe will transport the activated isotope stochastically thereby diminishing any localization correlations with the de-excitation gammas. A veto of the entire TPC is therefore necessary to conservatively remove \xe{137} signals.

\begin{figure}[h]
    \centering
    \includegraphics[width=\linewidth]{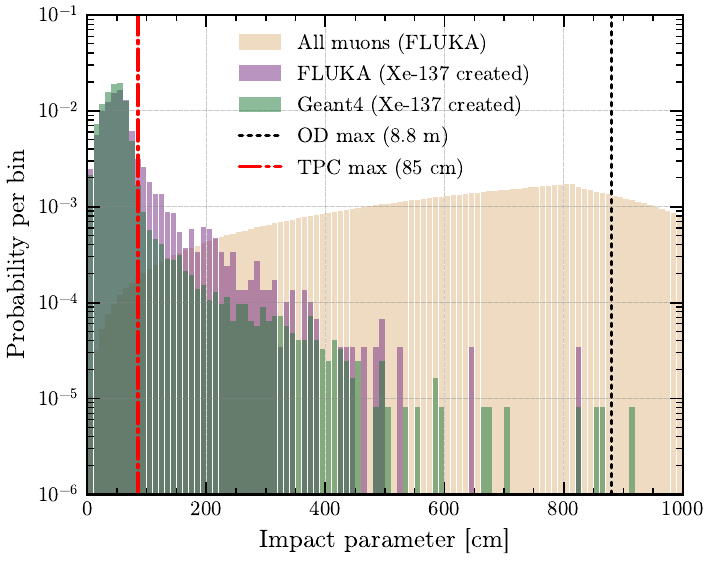}
    \caption{Normalized muon impact parameter distributions from both \code{Geant4} and \code{FLUKA} show that muons that pass near TPC contribute overwhelmingly to the production of ROCs. The orange distribution is that of the impact parameters of all muons simulated with FLUKA. The maximum impact parameters for muons passing the TPC and OD are shown by the red and black vertical lines, respectively}
    \label{fig:impacts}
\end{figure}

For each muon simulated, its distance of closest approach to the centre of the TPC is recorded. We call this number the \textit{impact parameter}. The distribution of impact parameters in our simulation is shown in Figure~\ref{fig:impacts}.

\begin{figure}[h]
    \centering
    \includegraphics[width=\linewidth]{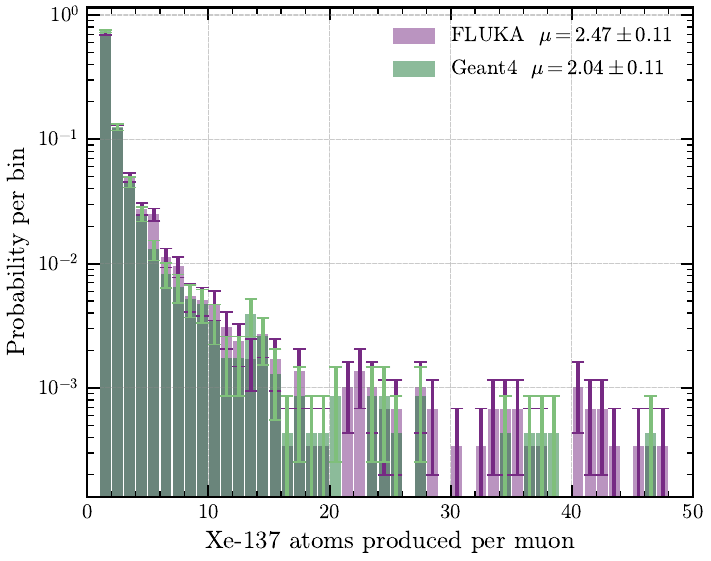}
    \caption{$^{137}$Xe multiplicity distributions from both \code{Geant4} and \code{FLUKA} excluding the zeroth bin. The mean multiplicities are presented in the legend. The median for both histograms is 1.0 (excluding 0). }
    \label{fig:multiplicities}
\end{figure}

Simulations show that $>$ \SI{99}{\percent} of muons producing ROC have impact parameters $\leq$ \SI{3.5}{\meter} as shown in Figure \ref{fig:impacts}. The target volume, with a radius of $\approx$ \SI{8}{\meter}, rejects a negligible number of background-inducing muons. The number of neutrons produced per muon (the neutron multiplicity) follows a decaying exponential distribution and, within SNOLAB's norite rock, a mean number of \SI{7.86}{\text{neutrons}\per\upmu} is expected \cite{MeiHime}. ROCs are therefore unlikely to be created independently but instead are likely produced in a series of activation events correlated in both space and time (see the \xe{137} multiplicity in Figure \ref{fig:multiplicities}). This combinatorial effect lends itself to developing a refined muon veto.

Considering that ROCs are produced through neutron capture $(\mathrm{n},\upgamma)$ interactions, events caused by ROCs may be vetoed by tagging their de-excitation gamma cascades.
A custom \code{Geant4} $\upgamma$-ray cascade model for ROCs was developed and validated for EXO-200; we use it here to evaluate the ability of nEXO to discern these signals within the TPC. In the following section, we discuss a light-only veto scheme using $^{137}$Xe as a case study. A combined charge and light signal analysis would provide the most information for standard analysis. However, here we focus exclusively on the scintillation light channel in this work owing to the difficulty of matching charge and light signals from multiple consecutive gamma cascades from neutron activation.

The ROCs were simulated in their appropriate locations as generated with \code{Shielding}. The scintillation signals from de-excitation $\upgamma$ cascades were simulated with a customized version of NEST~\cite{szydagis2011nest}, which simulates the production of ionization and scintillation signals in liquid xenon. Using nEXO's event reconstruction framework \cite{adhikari2021nexo}, we then create a scintillation-only reconstructed energy to tag the $\upgamma$ cascades. 

\begin{figure}[h]
    \centering
    \includegraphics[width=\linewidth]{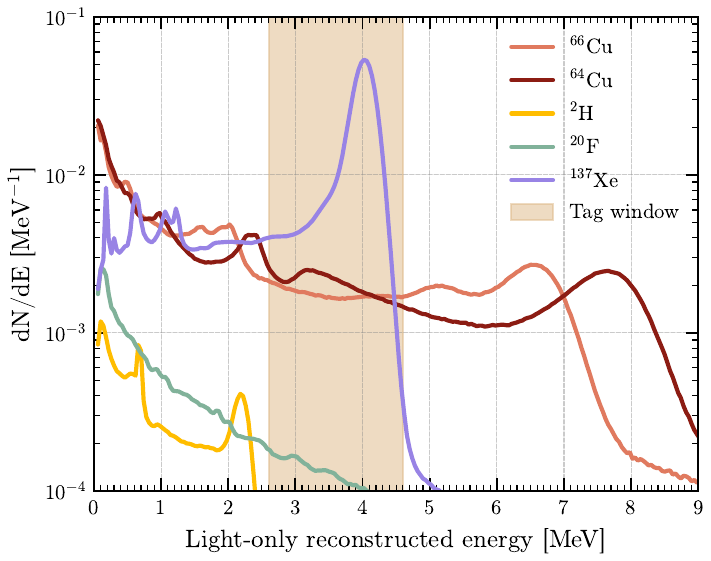}
    \caption{Post-neutron capture $\upgamma$-ray de-excitation energies as reconstructed only from light yields in the TPC. The post $(\mathrm{n},\upgamma)$ isotopes are listed in the legend. The tagging window, $[2.6,4.6]$ MeV, is highlighted in beige. Figure adapted from \cite{osti_1975602}}
    \label{fig:light_only}
\end{figure}

The light-only reconstructed energy spectra in nEXO for all nuclides for which the $\upgamma$ de-excitation cascades following neutron captures overlap that of \xe{137} are shown in Figure \ref{fig:light_only}. The energy range of 2.6 to 4.6 MeV is dominated by photons from $^{137}$Xe allowing highly efficient identification. The lower bound is above the H$(n,\upgamma)$D $\sim$ \SI{2.2}{\mega\electronvolt} $\upgamma$ energy, and excludes the region where accidental triggers due to Cu emissions would be most likely. For equivalent reasoning, the upper bound is set below the summed cascade energy of neutron capture on copper isotopes of $\sim 8$ MeV.

The efficiency of tagging a single \xe{137} by detecting energy deposits in the fiducial volume, within the tagging window is, conservatively, $p =$ \SI{70}{\percent} \cite{adhikari2021nexo}. The tagging efficiency, $\epsilon$, for $N$ nuclei produced by the same muon, is given by:
\begin{equation}
    \label{eqn:veto_opening_efficiency}
    \epsilon = 1 - (1 - p)^N ,
\end{equation}
which is evidently higher with the average multiplicity $N$ for $^{137}$Xe being $\bar{N} =$ \SI{2.26(0.08)} (Figure \ref{fig:multiplicities}). Taking the combined nonzero multiplicity from \code{FLUKA} and \code{Geant4}, the mean \xe{137} tagging efficiency is $\bar{\epsilon} =$ \SI{93.4(4.0)}{\percent}. 

We assume no event pileup. That is, we neglect the possibility of more than one gamma cascade occurring at once as a consequence of two simultaneous neutron captures. With SiPM readout that is fast (\SI{1}{\micro\second}) compared to stochastic neutron thermalization with times ($\gtrsim$ \SI{100}{\micro\second}), it is unlikely that several $\upgamma$ cascades from $^{137}$Xe*  occur simultaneously.

The overall expected veto efficiency is:
\begin{equation}
    \label{eqn:rejection_efficiency}
    \epsilon_{\text{X}}(\tau) = \bar{\epsilon}\left[1 - \left(\frac{1}{2}\right)^{\frac{\tau}{T_{1/2}}}\right] ,
\end{equation}
where $\tau$ is the time for which the datastream will be vetoed. For \xe{137}, $T_{1/2} \approx $ \SI{3.8}{\minute}, and choosing $\tau =$ \SI{25}{\min} to exceed 7 half-lives, the veto efficiency is $\epsilon_\text{X} =$ \SI{92.4(4.0)}{\percent}.



\subsection{Uncertainty Estimation for \xe{137} Background}

A muon-level bootstrap procedure was developed to obtain a representative estimate of the variance in ROC production and to quantify the residual, untagged \xe{137} population after implementation of the light-only veto scheme described above.

For both \code{Geant4} and \code{FLUKA}, each simulated muon was stored together with its complete event cascade and all ROCs produced within that cascade. These muon-level records were subsequently resampled with replacement to construct independent realizations corresponding to one decade of exposure, with each realization containing \(>3\times10^5\) muons. Because the complete cascade is retained as the resampling unit, correlations among ROCs produced by a common muon shower are preserved.

\begin{figure}[h]
    \centering
    \includegraphics[width=\linewidth]{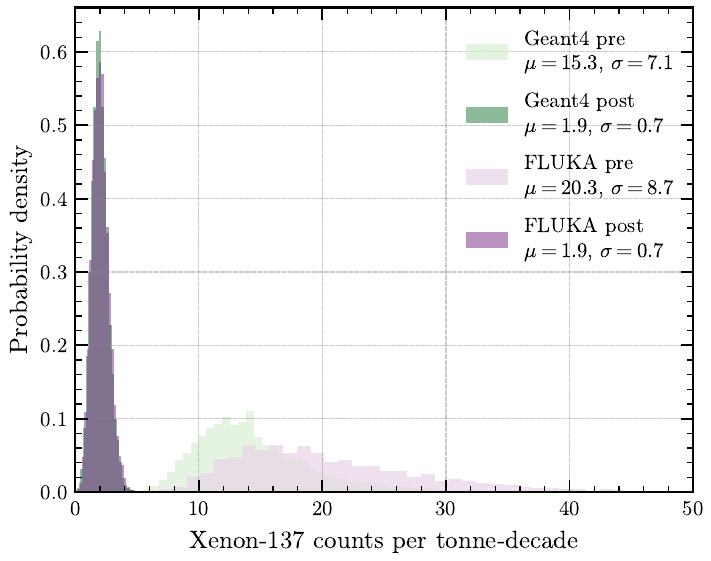}
    \caption{Estimates of probability density functions for the production rate of $^{137}$Xe atoms in the full nEXO LXe volume (pre-veto), and for those not removed by the light-only veto (post-veto). These estimates were obtained with the aforementioned bootstrap procedure.}
    \label{fig:pre-post-veto}
\end{figure}

For each decade, the ROCs produced were tallied. For each $^{137}$Xe atom produced, a random number $\xi \in [0,1)$ was sampled. If $\xi < R_X$, the atom was removed from the dataset to simulate the response of the light-only veto. This procedure was repeated for all $^{137}$Xe atoms in each bootstrap realization to obtain the corresponding \textit{post-veto} populations. The resulting pre- and post-veto $^{137}$Xe abundance distributions are shown in Figure~\ref{fig:pre-post-veto}.

\subsection{Deadtime Impact}
\begin{figure}[h]
    \centering
    \includegraphics[width=\linewidth]{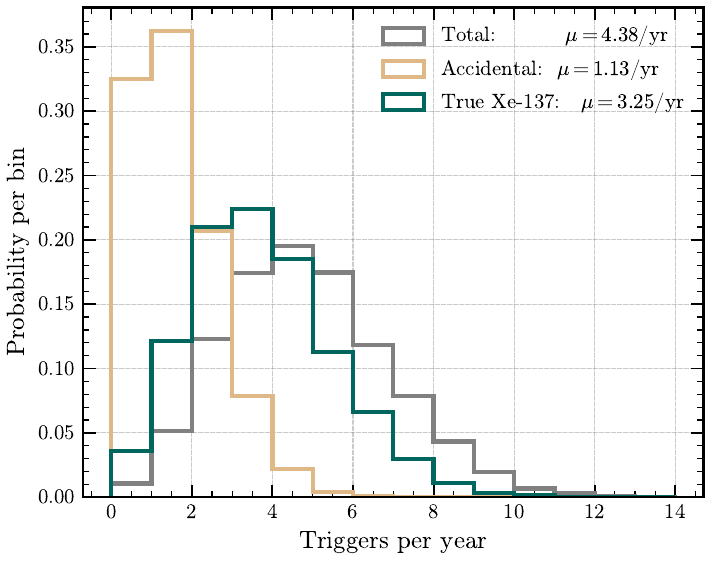}
    \caption{True, accidental, and total trigger counts per year estimated using the light-only de-excitation $\upgamma$ cascade veto scheme described herein.}
    \label{fig:triggered}
\end{figure}

Every veto window opened diminishes nEXO's sensitivity reach by reducing its exposure time. We estimate the anticipated \textit{deadtime}, due to both genuine $^{137}$Xe events and false coincidences; the distributions of anticipated deadtime triggers are shown in Figure \ref{fig:triggered}. The deadtime due to false coincidences was estimated from the likelihood of $^{63,65}$Cu activation to trigger a veto window of length $\tau$ for a light-only energy in the range $[2.6,4.6]$ MeV deposited within a time window of $[0,10]$ ms following a muon passage. 

The false-coincidence deadtime is determined using the same prescription described above for $^{137}$Xe but for muons producing $^{63,65}$Cu and no $^{137}$Xe. For the copper isotopes, the probability of opening a veto window is $p < 10\%$ proportional to the area underneath the respective curves in Figure \ref{fig:light_only} within the energy window.

The total fraction of false tags is $\sim9\%$, resulting in the average amount of deadtime due to false coincidences being $\approx$ \SI{30}{\minute\per\year}. Although there may be contributions due to de-excitations following neutron captures on other species, their contribution to the total deadtime is estimated to be small due to both the production rates being small, and, the energy window excluding most of their reconstructed energy spectra. The deadtime accrued only from true $^{136}$Xe$(\mathrm{n},\upgamma)$ tags has a marginal effect amounting to a mere $\sim$80 minutes per year. The total deadtime averages to approximately 2 hours per year.

\subsubsection*{Impact on sensitivity to \nubb{}}

\begin{figure}[h]
    \centering
    \includegraphics[width=\linewidth]{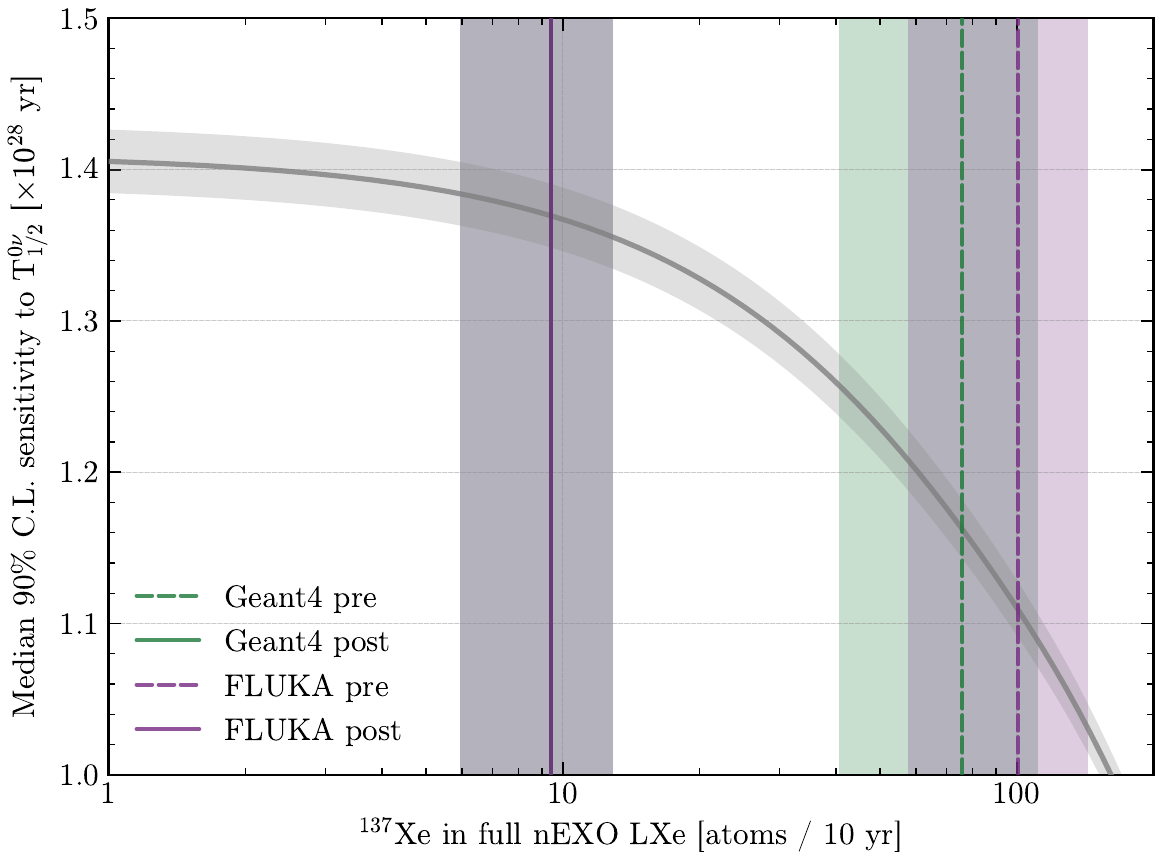}
    \caption{Median nEXO sensitivity to the \nubb{} half life as a function of the number of $^{137}$Xe atoms in the full LXe volume shown pre- and post-veto. Data are rescaled from \cite{adhikari2021nexo} to obtain appropriate x-axis units. The grey curve indicates the median exclusion sensitivity from 5000 simulated datasets for a given $^{137}$Xe background level, while the band indicates $1 \sigma$ coverage obtained by bootstrapping 50\% of the datasets and recalculating the median in 20,000 trials. 
    The x-axis is $^{137}$Xe atoms produced within the nEXO LXe; only a small fraction $\lesssim 1\%$ will remain in the dataset after energy, standoff, and topological cuts.}
    \label{fig:sensitivity_scaled}
\end{figure}

Using the sensitivity-vs-$^{137}$Xe-rate scaling relation calculated for nEXO in Ref~\cite{adhikari2021nexo}, reproduced here in Figure~\ref{fig:sensitivity_scaled}, we can quantify the impact of this veto on the sensitivity of nEXO to \ndbd{}. We find that the veto improves the sensitivity by $\sim$20\%, with only minor differences in the results from the \code{Geant4} and \code{FLUKA} simulations, and enables nEXO to achieve the projected half-life sensitivity of \SI{1.35(2)e28}{\year} reported in Ref.~\cite{adhikari2021nexo}.


\section{Conclusion}

We present a detailed evaluation of nEXO's cosmogenic backgrounds at SNOLAB as evaluated using recent versions of the \code{FLUKA} and \code{Geant4} codes with standardized features, physics lists and a custom muon generator. The independently run simulations yield results consistent within uncertainties, with \code{FLUKA} finding slightly higher activation rates.

A scintillation light-only TPC veto scheme is proposed to mitigate the negative effect of ROCs on nEXO sensitivity. In particular, we estimate the efficiency with which $^{137}$Xe events could be removed from the dataset using only the TPC response from $(\mathrm{n},\upgamma)$ de-excitations resulting from their activation. We predict a rejection efficiency of $(92.4 \pm 4.0)\%$ using this veto scheme.

This work supports the previously published assumptions in \cite{adhikari2021nexo} on the cosmogenic backgrounds to nEXO, post-veto. Consequently we report no change to the forecasted sensitivity of nEXO.

Future studies may investigate the tagging of individual de-excitation lines in the charge channel, thereby improving the overall veto efficiency and reducing deadtime accumulation. Furthermore, we did not discuss the role of nEXO's Outer Detector in tagging through-going muons via their produced Cherenkov light. This system, and its optimization will be addressed in a future publication.

\begin{acknowledgments}
This work was supported in part by Laboratory Directed Research and Development (LDRD) programs at Brookhaven National Laboratory (BNL), Lawrence Livermore National Laboratory (LLNL), Pacific Northwest National Laboratory (PNNL), and SLAC National Accelerator Laboratory. The authors gratefully acknowledge support for nEXO from the Office of Nuclear Physics within DOE’s Office of Science under grants/contracts DE-AC02-76SF00515, DE-FG02-01ER41166, DE-SC002305, DE-FG02-93ER40789, DE-SC0021388, DE-SC0012704, DE-AC52-07NA27344, DE‐SC0017970, DE-AC05-76RL01830, DE-SC0012654, DE-SC0021383, DE-SC0014517, DE-SC0024666, DE-SC0020509, DE-SC0024677 and support by the US National Science Foundation grants NSF PHY-2111213 and NSF PHY-2011948; from NSERC SAPPJ-2022-00021, CFI 39881, FRQNT 2019-NC-255821, and the CFREF Arthur B. McDonald Canadian Astroparticle Physics Research Institute in Canada; from IBS-R016-D1 in South Korea; and from CAS in China.
\end{acknowledgments}

\bibliographystyle{apsrev4-2}
\bibliography{bibliography}

\onecolumn
\section{Appendix}
\begin{table*}[h]
    \centering
    \caption{Summary of important options enabled for \code{FLUKA} simulations.}
    \label{tab:fluka_physics}
    \begin{tabular}{|c|c|c|}
    \multicolumn{3}{c}{Physics Options}\\
    \hline
    Category & Option (if relevant) & Description \\
    \hline
    DEFAULTS & PRECISIO & Default mode for precision simulations\\ 
    \hline
    MUPHOTON &  & Activates muon-nuclear interactions \\
    \hline
    PHOTONUC &  & Activates photonuclear interactions\\
    \hline
    PHOTONUC & MUMUPAIR & Activates muon pair production \\
    \hline
    PAIRBREM & Both & Activates pair production and bremsstrahlung \\
    \hline
    PHYSICS  & COALESCE & Activates the coalescence mechanism \\
    \hline
    PHYSICS  & EVAPORAT & Activates the emission of heavy fragments \\
    \hline
    PHYSICS  & EM-DISSO & Activates the electromagnetic dissociation of ions \\
    \hline
    \multirow{2}{*}{LOW-PWXS} & \multirow{2}{*}{JEFF-3.3} & Links JEFF $\sigma$ database for neutron interactions \\
     & & Activates point-wise treatment for neutron transport \\
     \hline
     \multicolumn{3}{l}{}\\
 
    \multicolumn{3}{c}{Compilation Options}\\
    \hline
    Feature & \multicolumn{2}{|c|}{Description} \\
    \hline
   DPMJET & \multicolumn{2}{|c|}{Event generator activated for E $>$ 5 GeV per nucleon} \\
   RQMD & \multicolumn{2}{|c|}{Event generator activated for E $>$ 125 keV per nucleon} \\
   \hline
    \end{tabular}
\end{table*}

\textbf{Isotopes Produced in Geant4 and FLUKA Simulations}

\begin{longtable}{@{}lllllll@{}}
\caption{Full isotope production rates from \code{Geant4} and \code{FLUKA} with combined statistical and systematic uncertainties, defined as above.} \label{tab:full-isotope-rates} \\
\toprule
Nuclide & Half-life & Decay energy (keV) & FLUKA yearly & FLUKA error & G4 yearly & G4 error \\
\midrule
\endfirsthead
\caption[]{Full isotope production rates from Geant4 and FLUKA.} \\
\toprule
Nuclide & Half-life & Decay energy (keV) & FLUKA yearly & FLUKA error & G4 yearly & G4 error \\
\midrule
\endhead
\midrule
\multicolumn{7}{r}{Continued on next page} \\
\midrule
\endfoot
\bottomrule
\endlastfoot
$^{1}\mathrm{H}$ & stable & --- & 1.84e+02 & 5.64e+00 & 2.88e+04 & 1.50e+03 \\
$^{19}\mathrm{F}$ & stable & --- & 2.94e+02 & 1.10e+01 & 1.37e+04 & 7.00e+02 \\
$^{12}\mathrm{C}$ & stable & --- & 2.10e+01 & 8.01e-01 & 9.26e+03 & 4.70e+02 \\
$^{136}\mathrm{Xe}$ & 3.6e+20 y & \qbb{} & 5.43e+01 & 2.04e+00 & 4.34e+03 & 2.20e+02 \\
$^{16}\mathrm{O}$ & stable & --- & 2.47e+01 & 9.39e-01 & 2.00e+03 & 1.00e+02 \\
$^{2}\mathrm{H}$ & stable & --- & 2.92e+02 & 1.07e+01 & 3.01e+02 & 1.50e+01 \\
$^{134}\mathrm{Xe}$ & stable & --- & 2.50e+01 & 9.51e-01 & 4.25e+02 & 2.20e+01 \\
$^{4}\mathrm{He}$ & stable & --- & 2.17e+02 & 7.83e+00 & 1.65e+02 & 8.40e+00 \\
$^{63}\mathrm{Cu}$ & stable & --- & 8.99e+00 & 3.46e-01 & 1.76e+02 & 9.00e+00 \\
$^{65}\mathrm{Cu}$ & stable & --- & 3.05e+00 & 1.28e-01 & 1.26e+02 & 6.40e+00 \\
$^{13}\mathrm{C}$ & stable & --- & 1.61e+01 & 6.20e-01 & 1.12e+02 & 5.70e+00 \\
$^{15}\mathrm{N}$ & stable & --- & 3.99e+01 & 1.50e+00 & 5.09e+01 & 2.60e+00 \\
$^{64}\mathrm{Cu}$ & 12.7 h & --- & 4.91e+01 & 1.73e+00 & 2.11e+01 & 1.10e+00 \\
$^{18}\mathrm{O}$ & stable & --- & 3.66e+01 & 1.38e+00 & 3.02e+01 & 1.60e+00 \\
$^{135}\mathrm{Xe}$ & 9.14 h & --- & 3.32e+01 & 1.26e+00 & 2.16e+01 & 1.10e+00 \\
$^{17}\mathrm{O}$ & stable & --- & 2.38e+01 & 9.07e-01 & 2.53e+01 & 1.30e+00 \\
$^{11}\mathrm{B}$ & stable & --- & 3.29e+01 & 1.25e+00 & 1.07e+01 & 5.70e-01 \\
$^{20}\mathrm{F}$ & 11.1 s & 7.02e+03 & 1.22e+01 & 4.75e-01 & 3.00e+01 & 1.60e+00 \\
$^{18}\mathrm{F}$ & 1.83 h & --- & 2.05e+01 & 7.82e-01 & 1.53e+01 & 8.00e-01 \\
$^{135}\mathrm{Cs}$ & 2.3e+06 y & --- & 3.36e+01 & 1.27e+00 & 7.49e-02 & 1.40e-02 \\
$^{3}\mathrm{H}$ & 12.3 y & --- & 1.92e+01 & 6.69e-01 & 1.38e+01 & 7.30e-01 \\
$^{64}\mathrm{Ni}$ & stable & --- & 3.04e+01 & 1.07e+00 & 3.36e-02 & 9.50e-03 \\
$^{137}\mathrm{Xe}$ & 3.82 min & 4.16e+03 & 1.04e+01 & 4.05e-01 & 1.48e+01 & 7.80e-01 \\
$^{14}\mathrm{N}$ & stable & --- & 1.32e+01 & 5.12e-01 & 1.04e+01 & 5.60e-01 \\
$^{28}\mathrm{Si}$ & stable & --- & 1.58e-02 & 4.78e-03 & 2.13e+01 & 1.10e+00 \\
$^{64}\mathrm{Zn}$ & stable & --- & 1.89e+01 & 6.78e-01 & 2.58e-03 & 1.13e-02 \\
$^{16}\mathrm{N}$ & 7.13 s & 3.73e+06 & 7.55e+00 & 3.00e-01 & 1.11e+01 & 5.90e-01 \\
$^{8}\mathrm{Be}$ & 7.02e-08 ns & 3.73e+06 & 1.66e-16 & 4.88e-10 & 1.82e+01 & 9.50e-01 \\
$^{11}\mathrm{C}$ & 20.4 min & --- & 1.08e+01 & 4.22e-01 & 5.63e+00 & 3.10e-01 \\
$^{3}\mathrm{He}$ & stable & --- & 1.18e+01 & 4.42e-01 & 3.73e+00 & 2.10e-01 \\
$^{66}\mathrm{Cu}$ & 5.12 min & 2.64e+03 & 1.01e+01 & 3.73e-01 & 4.66e+00 & 2.60e-01 \\
$^{59}\mathrm{Ni}$ & 7.6e+04 y & --- & 1.44e+01 & 5.51e-01 & 8.27e-02 & 1.50e-02 \\
$^{14}\mathrm{C}$ & 5.7e+03 y & --- & 6.14e+00 & 2.47e-01 & 7.43e+00 & 4.00e-01 \\
$^{20}\mathrm{Ne}$ & stable & --- & 1.23e+01 & 4.76e-01 & 4.91e-02 & 1.20e-02 \\
$^{7}\mathrm{Li}$ & stable & --- & 1.06e+01 & 4.10e-01 & 9.15e-01 & 6.70e-02 \\
$^{133}\mathrm{Xe}$ & 5.25 d & --- & 5.66e+00 & 2.29e-01 & 5.12e+00 & 2.80e-01 \\
$^{137}\mathrm{Cs}$ & 30.1 y & --- & 1.04e+01 & 4.05e-01 & 2.58e-03 & 1.13e-02 \\
$^{66}\mathrm{Zn}$ & stable & --- & 1.01e+01 & 3.73e-01 & 2.58e-03 & 1.13e-02 \\
$^{6}\mathrm{Li}$ & stable & --- & 9.54e+00 & 3.70e-01 & 5.04e-01 & 4.40e-02 \\
$^{9}\mathrm{Be}$ & stable & --- & 5.81e+00 & 2.34e-01 & 3.84e+00 & 2.20e-01 \\
$^{10}\mathrm{B}$ & stable & --- & 4.81e+00 & 1.97e-01 & 4.83e+00 & 2.70e-01 \\
$^{57}\mathrm{Co}$ & 272 d & --- & 7.84e+00 & 3.08e-01 & 3.62e-02 & 9.80e-03 \\
$^{132}\mathrm{Xe}$ & stable & --- & 3.29e+00 & 1.40e-01 & 3.69e+00 & 2.10e-01 \\
$^{133}\mathrm{Cs}$ & stable & --- & 6.18e+00 & 2.49e-01 & 9.56e-02 & 1.60e-02 \\
$^{57}\mathrm{Fe}$ & stable & --- & 4.99e+00 & 2.01e-01 & 1.55e-02 & 6.40e-03 \\
$^{15}\mathrm{O}$ & 2.04 min & 2.75e+03 & 1.85e+00 & 8.61e-02 & 2.87e+00 & 1.70e-01 \\
$^{19}\mathrm{O}$ & 26.9 s & 4.82e+03 & 1.71e+00 & 8.07e-02 & 2.85e+00 & 1.70e-01 \\
$^{62}\mathrm{Ni}$ & stable & --- & 3.88e+00 & 1.40e-01 & 6.23e-01 & 5.10e-02 \\
$^{61}\mathrm{Ni}$ & stable & --- & 4.13e+00 & 1.49e-01 & 1.73e-01 & 2.30e-02 \\
$^{7}\mathrm{Be}$ & 53.2 d & --- & 3.25e+00 & 1.39e-01 & 5.87e-01 & 4.90e-02 \\
$^{10}\mathrm{Be}$ & 1.51e+06 y & --- & 1.91e+00 & 8.85e-02 & 1.57e+00 & 1.00e-01 \\
$^{58}\mathrm{Ni}$ & stable & --- & 3.22e+00 & 1.36e-01 & 6.20e-02 & 1.30e-02 \\
$^{131}\mathrm{Xe}$ & stable & --- & 1.56e+00 & 7.50e-02 & 1.49e+00 & 9.80e-02 \\
$^{27}\mathrm{Al}$ & stable & --- & 1.29e-02 & 4.32e-03 & 2.97e+00 & 1.70e-01 \\
$^{62}\mathrm{Cu}$ & 9.67 min & 3.96e+03 & 2.47e+00 & 1.04e-01 & 4.08e-01 & 3.90e-02 \\
$^{130}\mathrm{Xe}$ & stable & --- & 1.32e+00 & 6.56e-02 & 1.47e+00 & 9.70e-02 \\
$^{63}\mathrm{Ni}$ & 101 y & --- & 2.44e+00 & 1.01e-01 & 1.21e-01 & 1.90e-02 \\
$^{56}\mathrm{Fe}$ & stable & --- & 2.49e+00 & 1.06e-01 & 2.07e-02 & 7.40e-03 \\
$^{60}\mathrm{Ni}$ & stable & --- & 2.15e+00 & 8.74e-02 & 1.40e-01 & 2.00e-02 \\
$^{17}\mathrm{F}$ & 1.07 min & 2.76e+03 & 1.30e+00 & 6.48e-02 & 6.61e-01 & 5.30e-02 \\
$^{12}\mathrm{B}$ & 20.2 ms & 1.34e+04 & 1.18e+00 & 6.03e-02 & 5.87e-01 & 4.90e-02 \\
$^{58}\mathrm{Fe}$ & stable & --- & 1.74e+00 & 7.78e-02 & 2.07e-02 & 7.40e-03 \\
$^{58}\mathrm{Co}$ & 70.9 d & 2.31e+03 & 1.70e+00 & 7.66e-02 & 2.33e-02 & 7.80e-03 \\
$^{17}\mathrm{N}$ & 4.17 s & 9.44e+05 & 1.70e-01 & 1.69e-02 & 1.33e+00 & 9.00e-02 \\
$^{128}\mathrm{Xe}$ & stable & --- & 7.14e-01 & 4.16e-02 & 7.78e-01 & 6.00e-02 \\
$^{8}\mathrm{Li}$ & 840 ms & 1.60e+04 & 1.23e+00 & 6.20e-02 & 1.11e-01 & 1.80e-02 \\
$^{129}\mathrm{Xe}$ & stable & --- & 5.18e-01 & 3.34e-02 & 7.73e-01 & 5.90e-02 \\
$^{29}\mathrm{Si}$ & stable & --- & 1.43e-02 & 4.56e-03 & 1.20e+00 & 8.20e-02 \\
$^{13}\mathrm{N}$ & 9.96 min & 2.22e+03 & 5.77e-01 & 3.59e-02 & 5.30e-01 & 4.60e-02 \\
$^{59}\mathrm{Co}$ & stable & --- & 1.07e+00 & 4.98e-02 & 2.84e-02 & 8.70e-03 \\
$^{56}\mathrm{Co}$ & 77.2 d & 4.57e+03 & 1.07e+00 & 5.54e-02 & 5.17e-03 & 1.53e-02 \\
$^{55}\mathrm{Fe}$ & 2.74 y & --- & 9.95e-01 & 5.16e-02 & 2.58e-02 & 8.30e-03 \\
$^{54}\mathrm{Fe}$ & stable & --- & 1.00e+00 & 5.25e-02 & 1.03e-02 & 2.22e-02 \\
$^{57}\mathrm{Ni}$ & 1.48 d & 3.26e+03 & 9.55e-01 & 5.13e-02 & 1.81e-02 & 6.90e-03 \\
$^{6}\mathrm{He}$ & 807 ms & 3.51e+03 & 8.42e-01 & 4.62e-02 & 1.14e-01 & 1.80e-02 \\
$^{131}\mathrm{I}$ & 8.03 d & --- & 4.32e-01 & 2.96e-02 & 4.01e-01 & 3.80e-02 \\
$^{127}\mathrm{I}$ & stable & --- & 5.31e-01 & 3.39e-02 & 2.84e-01 & 3.10e-02 \\
$^{133}\mathrm{I}$ & 20.8 h & --- & 4.06e-01 & 2.85e-02 & 3.98e-01 & 3.80e-02 \\
$^{129}\mathrm{I}$ & 1.57e+07 y & --- & 4.22e-01 & 2.92e-02 & 3.33e-01 & 3.40e-02 \\
$^{30}\mathrm{Si}$ & stable & --- & 1.29e-02 & 4.32e-03 & 6.98e-01 & 5.50e-02 \\
$^{134}\mathrm{I}$ & 52.5 min & 4.08e+03 & 3.67e-01 & 2.67e-02 & 3.31e-01 & 3.40e-02 \\
$^{127}\mathrm{Xe}$ & 36.3 d & --- & 2.24e-01 & 1.98e-02 & 4.60e-01 & 4.20e-02 \\
$^{10}\mathrm{C}$ & 19.3 s & 3.65e+03 & 2.49e-01 & 2.10e-02 & 4.06e-01 & 3.80e-02 \\
$^{126}\mathrm{Xe}$ & stable & --- & 2.88e-01 & 2.30e-02 & 3.39e-01 & 3.40e-02 \\
$^{135}\mathrm{I}$ & 6.58 h & 2.63e+03 & 2.86e-01 & 2.29e-02 & 3.26e-01 & 3.30e-02 \\
$^{51}\mathrm{V}$ & stable & --- & 3.00e-01 & 2.32e-02 & --- & --- \\
$^{132}\mathrm{I}$ & 2.29 h & 3.58e+03 & 3.28e-01 & 2.49e-02 & 2.71e-01 & 3.00e-02 \\
$^{128}\mathrm{I}$ & 25 min & --- & 3.16e-01 & 2.43e-02 & 2.56e-01 & 2.90e-02 \\
$^{130}\mathrm{I}$ & 12.4 h & 2.94e+03 & 3.19e-01 & 2.45e-02 & 2.48e-01 & 2.80e-02 \\
$^{9}\mathrm{B}$ & 6.81e-12 ns & 9.39e+05 & --- & --- & 2.79e-01 & 3.00e-02 \\
$^{53}\mathrm{Mn}$ & 3.74e+06 y & --- & 5.44e-01 & 3.38e-02 & 1.03e-02 & 2.22e-02 \\
$^{52}\mathrm{Cr}$ & stable & --- & 5.03e-01 & 3.17e-02 & 1.55e-02 & 6.40e-03 \\
$^{52}\mathrm{Mn}$ & 5.59 d & 4.71e+03 & 2.56e-01 & 2.11e-02 & --- & --- \\
$^{15}\mathrm{C}$ & 2.45 s & 9.77e+03 & 1.33e-01 & 1.47e-02 & 3.54e-01 & 3.50e-02 \\
$^{137}\mathrm{Ba}$ & stable & --- & 2.36e-01 & 2.04e-02 & --- & --- \\
$^{125}\mathrm{Te}$ & stable & --- & 3.45e-01 & 2.57e-02 & 1.14e-01 & 1.80e-02 \\
$^{54}\mathrm{Mn}$ & 312 d & --- & 2.29e-01 & 1.97e-02 & --- & --- \\
$^{125}\mathrm{I}$ & 59.4 d & --- & 2.55e-01 & 2.14e-02 & 1.94e-01 & 2.40e-02 \\
$^{61}\mathrm{Cu}$ & 3.34 h & 2.24e+03 & 2.68e-01 & 2.15e-02 & 1.60e-01 & 2.20e-02 \\
$^{126}\mathrm{Te}$ & stable & --- & 2.11e-01 & 1.91e-02 & 1.83e-01 & 2.40e-02 \\
$^{124}\mathrm{Te}$ & stable & --- & 2.29e-01 & 2.00e-02 & 1.65e-01 & 2.20e-02 \\
$^{13}\mathrm{B}$ & 17.3 ms & 9.48e+05 & 2.88e-01 & 2.30e-02 & 1.06e-01 & 1.70e-02 \\
$^{126}\mathrm{I}$ & 12.9 d & --- & 1.86e-01 & 1.77e-02 & 1.96e-01 & 2.50e-02 \\
$^{55}\mathrm{Mn}$ & stable & --- & 3.34e-01 & 2.45e-02 & 5.17e-03 & 1.53e-02 \\
$^{122}\mathrm{Te}$ & stable & --- & 1.87e-01 & 1.78e-02 & 1.40e-01 & 2.00e-02 \\
$^{19}\mathrm{Ne}$ & 17.2 s & 3.24e+03 & 9.74e-02 & 1.24e-02 & 2.07e-01 & 2.50e-02 \\
$^{51}\mathrm{Cr}$ & 27.7 d & --- & 2.77e-01 & 2.20e-02 & 7.75e-03 & 1.92e-02 \\
$^{14}\mathrm{O}$ & 1.18 min & 5.14e+03 & 2.86e-02 & 6.49e-03 & 2.53e-01 & 2.90e-02 \\
$^{125}\mathrm{Xe}$ & 16.9 h & --- & 8.30e-02 & 1.13e-02 & 1.89e-01 & 2.40e-02 \\
$^{123}\mathrm{Te}$ & 9.2e+16 y & --- & 1.68e-01 & 1.67e-02 & 1.01e-01 & 1.70e-02 \\
$^{8}\mathrm{B}$ & 770 ms & 3.75e+06 & 2.19e-01 & 1.95e-02 & 3.36e-02 & 9.50e-03 \\
$^{9}\mathrm{Li}$ & 178 ms & 9.52e+05 & 1.92e-01 & 1.81e-02 & 5.43e-02 & 1.20e-02 \\
$^{124}\mathrm{Xe}$ & 1.6e+14 y & --- & 7.16e-02 & 1.05e-02 & 1.71e-01 & 2.30e-02 \\
$^{129}\mathrm{Te}$ & 1.16 h & --- & 5.01e-02 & 8.67e-03 & 1.83e-01 & 2.40e-02 \\
$^{121}\mathrm{Sb}$ & stable & --- & 1.58e-01 & 1.61e-02 & 6.72e-02 & 1.40e-02 \\
$^{128}\mathrm{Te}$ & stable & --- & 1.02e-01 & 1.27e-02 & 1.14e-01 & 1.80e-02 \\
$^{123}\mathrm{I}$ & 13.2 h & --- & 8.35e-02 & 1.14e-02 & 1.32e-01 & 2.00e-02 \\
$^{124}\mathrm{I}$ & 4.18 d & 3.16e+03 & 1.09e-01 & 1.31e-02 & 1.01e-01 & 1.70e-02 \\
$^{17}\mathrm{C}$ & 193 ms & 9.47e+05 & --- & --- & 1.03e-01 & 1.70e-02 \\
$^{120}\mathrm{Te}$ & stable & --- & 9.31e-02 & 1.21e-02 & 1.14e-01 & 1.80e-02 \\
$^{127}\mathrm{Te}$ & 9.35 h & --- & 9.45e-02 & 1.22e-02 & 1.11e-01 & 1.80e-02 \\
$^{50}\mathrm{Cr}$ & 1.3e+18 y & --- & 1.98e-01 & 1.82e-02 & 7.75e-03 & 1.92e-02 \\
$^{130}\mathrm{Te}$ & 7.9e+20 y & 2.53e+03 & 6.01e-02 & 9.55e-03 & 1.40e-01 & 2.00e-02 \\
$^{60}\mathrm{Co}$ & 5.27 y & 2.82e+03 & 1.68e-01 & 1.61e-02 & 3.10e-02 & 9.10e-03 \\
$^{49}\mathrm{V}$ & 330 d & --- & 1.90e-01 & 1.77e-02 & 7.75e-03 & 1.92e-02 \\
$^{47}\mathrm{Ti}$ & stable & --- & 9.88e-02 & 1.25e-02 & --- & --- \\
$^{16}\mathrm{C}$ & 747 ms & 9.45e+05 & 1.58e-02 & 4.78e-03 & 1.81e-01 & 2.30e-02 \\
$^{121}\mathrm{Te}$ & 19.2 d & --- & 1.15e-01 & 1.35e-02 & 7.75e-02 & 1.50e-02 \\
$^{48}\mathrm{Ti}$ & stable & --- & 1.82e-01 & 1.74e-02 & 1.03e-02 & 2.22e-02 \\
$^{53}\mathrm{Fe}$ & 8.51 min & 3.74e+03 & 9.31e-02 & 1.21e-02 & --- & --- \\
$^{10}\mathrm{Li}$ & stable & 9.40e+05 & --- & --- & 9.30e-02 & 1.60e-02 \\
$^{46}\mathrm{Ti}$ & stable & --- & 1.71e-01 & 1.66e-02 & 2.58e-03 & 1.13e-02 \\
$^{50}\mathrm{V}$ & 1.4e+17 y & --- & 8.02e-02 & 1.10e-02 & --- & --- \\
$^{18}\mathrm{N}$ & 624 ms & 3.74e+06 & --- & --- & 7.75e-02 & 1.50e-02 \\
$^{114}\mathrm{Sn}$ & stable & --- & 8.73e-02 & 1.16e-02 & 6.72e-02 & 1.40e-02 \\
$^{134}\mathrm{Cs}$ & 2.07 y & --- & 5.73e-02 & 9.30e-03 & 8.78e-02 & 1.60e-02 \\
$^{119}\mathrm{Te}$ & 16.1 h & 2.29e+03 & 6.00e-02 & 9.54e-03 & 7.75e-02 & 1.50e-02 \\
$^{121}\mathrm{I}$ & 2.12 h & 2.29e+03 & 6.73e-02 & 1.01e-02 & 6.98e-02 & 1.40e-02 \\
$^{14}\mathrm{B}$ & 12.5 ms & 2.06e+04 & --- & --- & 6.72e-02 & 1.40e-02 \\
$^{116}\mathrm{Sn}$ & stable & --- & 7.45e-02 & 1.07e-02 & 5.68e-02 & 1.20e-02 \\
$^{54}\mathrm{Cr}$ & stable & --- & 1.28e-01 & 1.42e-02 & 2.58e-03 & 1.13e-02 \\
$^{118}\mathrm{Sn}$ & stable & --- & 8.73e-02 & 1.16e-02 & 4.13e-02 & 1.10e-02 \\
$^{131}\mathrm{Cs}$ & 9.69 d & --- & 5.01e-02 & 8.67e-03 & 7.75e-02 & 1.50e-02 \\
$^{119}\mathrm{Sn}$ & stable & --- & 1.05e-01 & 1.28e-02 & 2.07e-02 & 7.40e-03 \\
$^{55}\mathrm{Co}$ & 17.5 h & 3.45e+03 & 1.20e-01 & 1.39e-02 & 5.17e-03 & 1.53e-02 \\
$^{48}\mathrm{V}$ & 16 d & 4.02e+03 & 1.19e-01 & 1.38e-02 & 5.17e-03 & 1.53e-02 \\
$^{4}\mathrm{H}$ & stable & 9.41e+05 & --- & --- & 6.20e-02 & 1.30e-02 \\
$^{122}\mathrm{I}$ & 3.63 min & 4.23e+03 & 6.16e-02 & 9.67e-03 & 5.94e-02 & 1.30e-02 \\
$^{136}\mathrm{I}$ & 1.39 min & 6.88e+03 & 2.58e-02 & 6.15e-03 & 9.04e-02 & 1.60e-02 \\
$^{123}\mathrm{Xe}$ & 2.08 h & 2.70e+03 & 3.87e-02 & 7.58e-03 & 7.75e-02 & 1.50e-02 \\
$^{132}\mathrm{Cs}$ & 6.48 d & --- & 5.30e-02 & 8.93e-03 & 6.20e-02 & 1.30e-02 \\
$^{65}\mathrm{Ni}$ & 2.52 h & --- & 5.73e-02 & 9.22e-03 & --- & --- \\
$^{129}\mathrm{Sb}$ & 4.37 h & 2.38e+03 & --- & --- & 5.43e-02 & 1.20e-02 \\
$^{131}\mathrm{Te}$ & 25 min & 2.23e+03 & 3.44e-02 & 7.13e-03 & 7.23e-02 & 1.40e-02 \\
$^{130}\mathrm{Cs}$ & 29.2 min & 2.98e+03 & 4.73e-02 & 8.41e-03 & 5.94e-02 & 1.30e-02 \\
$^{117}\mathrm{Sn}$ & stable & --- & 8.30e-02 & 1.13e-02 & 2.33e-02 & 7.80e-03 \\
$^{122}\mathrm{Xe}$ & 20.1 h & --- & 2.86e-02 & 6.49e-03 & 7.75e-02 & 1.50e-02 \\
$^{49}\mathrm{Ti}$ & stable & --- & 1.02e-01 & 1.25e-02 & 2.58e-03 & 1.13e-02 \\
$^{119}\mathrm{Sb}$ & 1.59 d & --- & 5.84e-02 & 9.40e-03 & 4.39e-02 & 1.10e-02 \\
$^{11}\mathrm{Be}$ & 13.8 s & 3.73e+06 & 9.02e-02 & 1.19e-02 & 1.03e-02 & 2.22e-02 \\
$^{15}\mathrm{B}$ & 9.93 ms & 1.91e+04 & --- & --- & 4.91e-02 & 1.20e-02 \\
$^{115}\mathrm{Sn}$ & stable & --- & 5.58e-02 & 9.18e-03 & 3.88e-02 & 1.00e-02 \\
$^{112}\mathrm{Sn}$ & stable & --- & 4.41e-02 & 8.12e-03 & 4.91e-02 & 1.20e-02 \\
$^{118}\mathrm{Te}$ & 6 d & --- & 3.58e-02 & 7.28e-03 & 5.68e-02 & 1.20e-02 \\
$^{129}\mathrm{Cs}$ & 1.34 d & --- & 2.72e-02 & 6.32e-03 & 6.46e-02 & 1.30e-02 \\
$^{132}\mathrm{Te}$ & 3.2 d & --- & 2.58e-02 & 6.15e-03 & 6.46e-02 & 1.30e-02 \\
$^{123}\mathrm{Sb}$ & stable & --- & 3.01e-02 & 6.66e-03 & 5.94e-02 & 1.30e-02 \\
$^{118}\mathrm{Sb}$ & 3.6 min & 3.66e+03 & 3.44e-02 & 7.13e-03 & 5.17e-02 & 1.20e-02 \\
$^{61}\mathrm{Co}$ & 1.65 h & --- & 6.73e-02 & 9.98e-03 & 1.81e-02 & 6.90e-03 \\
$^{45}\mathrm{Sc}$ & stable & --- & 8.02e-02 & 1.10e-02 & 5.17e-03 & 1.53e-02 \\
$^{5}\mathrm{Li}$ & 1.55e-14 ns & --- & 7.37e-25 & 3.25e-14 & 8.53e-02 & 1.50e-02 \\
$^{117}\mathrm{Sb}$ & 2.8 h & --- & 4.15e-02 & 7.86e-03 & 3.88e-02 & 1.00e-02 \\
$^{44}\mathrm{Ca}$ & stable & --- & 7.46e-02 & 1.07e-02 & 2.58e-03 & 1.13e-02 \\
$^{120}\mathrm{Sb}$ & 15.9 min & 2.68e+03 & 2.43e-02 & 5.97e-03 & 5.17e-02 & 1.20e-02 \\
$^{40}\mathrm{K}$ & 1.25e+09 y & --- & 3.72e-02 & 7.43e-03 & --- & --- \\
$^{113}\mathrm{Sn}$ & 115 d & --- & 2.86e-02 & 6.49e-03 & 4.39e-02 & 1.10e-02 \\
$^{28}\mathrm{Al}$ & 2.25 min & 4.64e+03 & --- & --- & 3.62e-02 & 9.80e-03 \\
$^{6}\mathrm{Be}$ & 1.16e-09 ns & 1.88e+06 & --- & --- & 3.62e-02 & 9.80e-03 \\
$^{134}\mathrm{Te}$ & 41.8 min & --- & --- & --- & 3.62e-02 & 9.80e-03 \\
$^{121}\mathrm{Xe}$ & 40.1 min & 3.77e+03 & --- & --- & 3.62e-02 & 9.80e-03 \\
$^{120}\mathrm{Sn}$ & stable & --- & 4.58e-02 & 8.28e-03 & 2.33e-02 & 7.80e-03 \\
$^{51}\mathrm{Mn}$ & 46.2 min & 3.21e+03 & 6.59e-02 & 1.00e-02 & 2.58e-03 & 1.13e-02 \\
$^{126}\mathrm{Cs}$ & 1.64 min & 4.80e+03 & --- & --- & 3.36e-02 & 9.50e-03 \\
$^{132}\mathrm{Sb}$ & 2.79 min & 5.55e+03 & --- & --- & 3.36e-02 & 9.50e-03 \\
$^{12}\mathrm{N}$ & 11 ms & 1.73e+04 & 2.15e-02 & 5.60e-03 & 4.39e-02 & 1.10e-02 \\
$^{49}\mathrm{Cr}$ & 42.3 min & 2.63e+03 & 5.87e-02 & 9.43e-03 & 5.17e-03 & 1.53e-02 \\
$^{113}\mathrm{In}$ & stable & --- & 4.69e-02 & 8.38e-03 & 1.55e-02 & 6.40e-03 \\
$^{25}\mathrm{Mg}$ & stable & --- & --- & --- & 3.10e-02 & 9.10e-03 \\
$^{18}\mathrm{Ne}$ & 1.67 s & 4.44e+03 & --- & --- & 3.10e-02 & 9.10e-03 \\
$^{126}\mathrm{Sb}$ & 12.3 d & 3.67e+03 & --- & --- & 3.10e-02 & 9.10e-03 \\
$^{131}\mathrm{Sb}$ & 23 min & 3.23e+03 & --- & --- & 3.10e-02 & 9.10e-03 \\
$^{44}\mathrm{Sc}$ & 3.97 h & 3.65e+03 & 5.87e-02 & 9.43e-03 & 2.58e-03 & 1.13e-02 \\
$^{111}\mathrm{Sn}$ & 35.3 min & 2.45e+03 & 2.15e-02 & 5.60e-03 & 3.88e-02 & 1.00e-02 \\
$^{116}\mathrm{Sb}$ & 15.8 min & 4.70e+03 & 3.44e-02 & 7.13e-03 & 2.58e-02 & 8.30e-03 \\
$^{124}\mathrm{Sb}$ & 60.2 d & 2.91e+03 & 2.58e-02 & 6.15e-03 & 3.36e-02 & 9.50e-03 \\
$^{111}\mathrm{Cd}$ & stable & --- & 5.30e-02 & 8.93e-03 & 5.17e-03 & 1.53e-02 \\
$^{111}\mathrm{In}$ & 2.8 d & --- & 3.99e-02 & 7.70e-03 & 1.81e-02 & 6.90e-03 \\
$^{16}\mathrm{B}$ & 1.26e-09 ns & 9.40e+05 & --- & --- & 2.84e-02 & 8.70e-03 \\
$^{26}\mathrm{Mg}$ & stable & --- & --- & --- & 2.84e-02 & 8.70e-03 \\
$^{128}\mathrm{Cs}$ & 3.64 min & 3.93e+03 & 2.15e-02 & 5.60e-03 & 3.36e-02 & 9.50e-03 \\
$^{41}\mathrm{Ca}$ & 9.94e+04 y & --- & 2.72e-02 & 6.32e-03 & --- & --- \\
$^{133}\mathrm{Te}$ & 12.5 min & 2.92e+03 & 1.86e-02 & 5.21e-03 & 3.36e-02 & 9.50e-03 \\
$^{125}\mathrm{Cs}$ & 46.7 min & 3.11e+03 & --- & --- & 2.58e-02 & 8.30e-03 \\
$^{16}\mathrm{F}$ & 5.04e-10 ns & 9.39e+05 & --- & --- & 2.58e-02 & 8.30e-03 \\
$^{115}\mathrm{Sb}$ & 32.1 min & 3.03e+03 & 2.72e-02 & 6.32e-03 & 2.33e-02 & 7.80e-03 \\
$^{109}\mathrm{Cd}$ & 1.26 y & --- & 2.43e-02 & 5.96e-03 & 2.58e-02 & 8.30e-03 \\
$^{122}\mathrm{Sb}$ & 2.72 d & --- & 1.29e-02 & 4.32e-03 & 3.62e-02 & 9.80e-03 \\
$^{125}\mathrm{Sb}$ & 2.76 y & --- & 1.72e-02 & 5.00e-03 & 3.10e-02 & 9.10e-03 \\
$^{114}\mathrm{Sb}$ & 3.49 min & 6.06e+03 & 2.72e-02 & 6.32e-03 & 2.07e-02 & 7.40e-03 \\
$^{116}\mathrm{Te}$ & 2.49 h & --- & 1.58e-02 & 4.78e-03 & 3.10e-02 & 9.10e-03 \\
$^{5}\mathrm{He}$ & 8.17e-15 ns & --- & --- & --- & 2.33e-02 & 7.80e-03 \\
$^{92}\mathrm{Mo}$ & stable & --- & --- & --- & 2.33e-02 & 7.80e-03 \\
$^{12}\mathrm{Be}$ & 21.3 ms & 1.17e+04 & 2.00e-02 & 5.41e-03 & 2.58e-02 & 8.30e-03 \\
$^{47}\mathrm{V}$ & 32.6 min & 2.93e+03 & 2.29e-02 & 5.79e-03 & --- & --- \\
$^{127}\mathrm{Cs}$ & 6.25 h & --- & 1.43e-02 & 4.56e-03 & 3.10e-02 & 9.10e-03 \\
$^{117}\mathrm{Te}$ & 1.03 h & 3.54e+03 & 1.15e-02 & 4.07e-03 & 3.36e-02 & 9.50e-03 \\
$^{110}\mathrm{Sn}$ & 4.15 h & --- & 1.15e-02 & 4.07e-03 & 3.10e-02 & 9.10e-03 \\
$^{104}\mathrm{Cd}$ & 57.7 min & --- & --- & --- & 2.07e-02 & 7.40e-03 \\
$^{105}\mathrm{Cd}$ & 55.5 min & 2.74e+03 & --- & --- & 2.07e-02 & 7.40e-03 \\
$^{7}\mathrm{He}$ & 1.89e-09 ns & --- & --- & --- & 2.07e-02 & 7.40e-03 \\
$^{130}\mathrm{Sb}$ & 39.5 min & 5.07e+03 & --- & --- & 2.07e-02 & 7.40e-03 \\
$^{109}\mathrm{Sn}$ & 18.1 min & 3.86e+03 & --- & --- & 2.07e-02 & 7.40e-03 \\
$^{120}\mathrm{Xe}$ & 40 min & --- & --- & --- & 2.07e-02 & 7.40e-03 \\
$^{102}\mathrm{Pd}$ & stable & --- & 2.58e-02 & 6.15e-03 & 1.55e-02 & 6.40e-03 \\
$^{53}\mathrm{Cr}$ & stable & --- & 3.87e-02 & 7.58e-03 & 2.58e-03 & 1.13e-02 \\
$^{106}\mathrm{Cd}$ & stable & --- & 2.00e-02 & 5.41e-03 & 2.07e-02 & 7.40e-03 \\
$^{8}\mathrm{He}$ & 119 ms & 9.48e+05 & 3.01e-02 & 6.66e-03 & 1.03e-02 & 2.22e-02 \\
$^{108}\mathrm{Cd}$ & stable & --- & 3.40e-02 & 7.09e-03 & 5.17e-03 & 1.53e-02 \\
$^{42}\mathrm{Ca}$ & stable & --- & 3.58e-02 & 7.28e-03 & 2.58e-03 & 1.13e-02 \\
$^{127}\mathrm{Sb}$ & 3.85 d & --- & 1.15e-02 & 4.07e-03 & 2.58e-02 & 8.30e-03 \\
$^{110}\mathrm{Cd}$ & stable & --- & 2.43e-02 & 5.97e-03 & 1.29e-02 & 5.80e-03 \\
$^{39}\mathrm{K}$ & stable & --- & 1.86e-02 & 5.21e-03 & --- & --- \\
$^{43}\mathrm{Ca}$ & stable & --- & 3.44e-02 & 7.13e-03 & 2.58e-03 & 1.13e-02 \\
$^{94}\mathrm{Mo}$ & stable & --- & --- & --- & 1.81e-02 & 6.90e-03 \\
$^{17}\mathrm{Ne}$ & 109 ms & 9.53e+05 & --- & --- & 1.81e-02 & 6.90e-03 \\
$^{133}\mathrm{Sb}$ & 2.34 min & 4.01e+03 & --- & --- & 1.81e-02 & 6.90e-03 \\
$^{121}\mathrm{Sn}$ & 1.13 d & --- & --- & --- & 1.81e-02 & 6.90e-03 \\
$^{114}\mathrm{Te}$ & 15.2 min & 2.61e+03 & --- & --- & 1.81e-02 & 6.90e-03 \\
$^{112}\mathrm{In}$ & 14.9 min & 2.58e+03 & 2.29e-02 & 5.79e-03 & 1.29e-02 & 5.80e-03 \\
$^{36}\mathrm{Cl}$ & 3.01e+05 y & --- & 3.29e-02 & 6.98e-03 & 2.58e-03 & 1.13e-02 \\
$^{120}\mathrm{I}$ & 1.36 h & 5.62e+03 & 1.72e-02 & 5.00e-03 & 1.81e-02 & 6.90e-03 \\
$^{136}\mathrm{Cs}$ & 13.2 d & 2.55e+03 & 1.43e-02 & 4.56e-03 & 2.07e-02 & 7.40e-03 \\
$^{119}\mathrm{I}$ & 19.1 min & 3.42e+03 & 2.15e-02 & 5.60e-03 & 1.29e-02 & 5.80e-03 \\
$^{46}\mathrm{Sc}$ & 83.8 d & 2.37e+03 & 3.15e-02 & 6.82e-03 & 2.58e-03 & 1.13e-02 \\
$^{9}\mathrm{C}$ & 126 ms & 3.74e+06 & 1.58e-02 & 4.78e-03 & 1.81e-02 & 6.90e-03 \\
$^{38}\mathrm{Ar}$ & stable & --- & 3.01e-02 & 6.66e-03 & 2.58e-03 & 1.13e-02 \\
$^{134}\mathrm{Ba}$ & stable & --- & 1.63e-02 & 4.87e-03 & --- & --- \\
$^{107}\mathrm{Ag}$ & stable & --- & 2.43e-02 & 5.97e-03 & 7.75e-03 & 1.92e-02 \\
$^{52}\mathrm{Fe}$ & 8.28 h & 2.38e+03 & 2.43e-02 & 5.97e-03 & 7.75e-03 & 1.92e-02 \\
$^{22}\mathrm{Na}$ & 2.6 y & 2.84e+03 & --- & --- & 1.55e-02 & 6.40e-03 \\
$^{100}\mathrm{Pd}$ & 3.63 d & --- & --- & --- & 1.55e-02 & 6.40e-03 \\
$^{94}\mathrm{Ru}$ & 51.8 min & --- & --- & --- & 1.55e-02 & 6.40e-03 \\
$^{96}\mathrm{Ru}$ & stable & --- & --- & --- & 1.55e-02 & 6.40e-03 \\
$^{113}\mathrm{Sb}$ & 6.67 min & 3.91e+03 & --- & --- & 1.55e-02 & 6.40e-03 \\
$^{128}\mathrm{Sb}$ & 9.05 h & 4.36e+03 & --- & --- & 1.55e-02 & 6.40e-03 \\
$^{123}\mathrm{Sn}$ & 129 d & --- & --- & --- & 1.55e-02 & 6.40e-03 \\
$^{63}\mathrm{Zn}$ & 38.5 min & 3.37e+03 & --- & --- & 1.55e-02 & 6.40e-03 \\
$^{90}\mathrm{Zr}$ & stable & --- & --- & --- & 1.55e-02 & 6.40e-03 \\
$^{24}\mathrm{Mg}$ & stable & --- & 1.29e-02 & 4.32e-03 & 1.81e-02 & 6.90e-03 \\
$^{60}\mathrm{Cu}$ & 23.7 min & 6.13e+03 & 2.72e-02 & 6.32e-03 & 2.58e-03 & 1.13e-02 \\
$^{56}\mathrm{Ni}$ & 6.08 d & --- & 2.72e-02 & 6.32e-03 & 2.58e-03 & 1.13e-02 \\
$^{104}\mathrm{Pd}$ & stable & --- & 2.72e-02 & 6.32e-03 & 2.58e-03 & 1.13e-02 \\
$^{103}\mathrm{Pd}$ & 17 d & --- & 1.15e-02 & 4.07e-03 & 1.81e-02 & 6.90e-03 \\
$^{136}\mathrm{Ba}$ & stable & --- & 1.43e-02 & 4.56e-03 & --- & --- \\
$^{43}\mathrm{Sc}$ & 3.89 h & 2.22e+03 & 1.43e-02 & 4.56e-03 & --- & --- \\
$^{34}\mathrm{S}$ & stable & --- & 2.58e-02 & 6.15e-03 & 2.58e-03 & 1.13e-02 \\
$^{45}\mathrm{Ti}$ & 3.08 h & --- & 2.29e-02 & 5.79e-03 & 5.17e-03 & 1.53e-02 \\
$^{85}\mathrm{Rb}$ & stable & --- & 1.40e-02 & 4.51e-03 & --- & --- \\
$^{108}\mathrm{In}$ & 58 min & 5.13e+03 & --- & --- & 1.29e-02 & 5.80e-03 \\
$^{108}\mathrm{Sn}$ & 10.3 min & --- & --- & --- & 1.29e-02 & 5.80e-03 \\
$^{82}\mathrm{Sr}$ & 25.4 d & --- & --- & --- & 1.29e-02 & 5.80e-03 \\
$^{115}\mathrm{Te}$ & 5.8 min & 4.94e+03 & --- & --- & 1.29e-02 & 5.80e-03 \\
$^{62}\mathrm{Zn}$ & 9.19 h & --- & --- & --- & 1.29e-02 & 5.80e-03 \\
$^{88}\mathrm{Zr}$ & 83.4 d & --- & --- & --- & 1.29e-02 & 5.80e-03 \\
$^{109}\mathrm{In}$ & 4.16 h & --- & 1.29e-02 & 4.32e-03 & 1.29e-02 & 5.80e-03 \\
$^{30}\mathrm{P}$ & 2.5 min & 4.23e+03 & 1.29e-02 & 4.32e-03 & --- & --- \\
$^{114}\mathrm{In}$ & 1.2 min & --- & 1.43e-02 & 4.56e-03 & 1.03e-02 & 2.22e-02 \\
$^{104}\mathrm{Ag}$ & 1.15 h & 4.28e+03 & 1.58e-02 & 4.78e-03 & 7.75e-03 & 1.92e-02 \\
$^{34}\mathrm{Cl}$ & 1.53 s & 5.49e+03 & 1.15e-02 & 4.07e-03 & --- & --- \\
$^{35}\mathrm{Cl}$ & stable & --- & 1.15e-02 & 4.07e-03 & --- & --- \\
$^{31}\mathrm{P}$ & stable & --- & 1.15e-02 & 4.07e-03 & --- & --- \\
$^{103}\mathrm{Rh}$ & stable & --- & 1.15e-02 & 4.07e-03 & --- & --- \\
$^{107}\mathrm{Cd}$ & 6.5 h & --- & 1.15e-02 & 4.07e-03 & 1.03e-02 & 2.22e-02 \\
$^{21}\mathrm{Ne}$ & stable & --- & 1.15e-02 & 4.07e-03 & 1.03e-02 & 2.22e-02 \\
$^{122}\mathrm{Sn}$ & stable & --- & 3.11e-04 & 6.67e-04 & 2.07e-02 & 7.40e-03 \\
$^{112}\mathrm{Cd}$ & stable & --- & 1.31e-02 & 4.37e-03 & 7.75e-03 & 1.92e-02 \\
$^{7}\mathrm{B}$ & 1.01e-08 ns & 9.41e+05 & --- & --- & 1.03e-02 & 2.22e-02 \\
$^{124}\mathrm{Cs}$ & 30.9 s & 5.93e+03 & --- & --- & 1.03e-02 & 2.22e-02 \\
$^{15}\mathrm{F}$ & 8.32e-09 ns & 9.40e+05 & --- & --- & 1.03e-02 & 2.22e-02 \\
$^{69}\mathrm{Ge}$ & 1.63 d & 2.23e+03 & --- & --- & 1.03e-02 & 2.22e-02 \\
$^{80}\mathrm{Kr}$ & stable & --- & --- & --- & 1.03e-02 & 2.22e-02 \\
$^{89}\mathrm{Mo}$ & 2.11 min & 5.61e+03 & --- & --- & 1.03e-02 & 2.22e-02 \\
$^{101}\mathrm{Pd}$ & 8.47 h & --- & --- & --- & 1.03e-02 & 2.22e-02 \\
$^{82}\mathrm{Rb}$ & 1.26 min & 4.40e+03 & --- & --- & 1.03e-02 & 2.22e-02 \\
$^{100}\mathrm{Ru}$ & stable & --- & --- & --- & 1.03e-02 & 2.22e-02 \\
$^{97}\mathrm{Ru}$ & 2.83 d & --- & --- & --- & 1.03e-02 & 2.22e-02 \\
$^{98}\mathrm{Ru}$ & stable & --- & --- & --- & 1.03e-02 & 2.22e-02 \\
$^{73}\mathrm{Se}$ & 7.15 h & 2.73e+03 & --- & --- & 1.03e-02 & 2.22e-02 \\
$^{126}\mathrm{Sn}$ & 2.3e+05 y & --- & --- & --- & 1.03e-02 & 2.22e-02 \\
$^{83}\mathrm{Sr}$ & 1.35 d & 2.27e+03 & --- & --- & 1.03e-02 & 2.22e-02 \\
$^{89}\mathrm{Zr}$ & 3.27 d & 2.83e+03 & --- & --- & 1.03e-02 & 2.22e-02 \\
$^{41}\mathrm{K}$ & stable & --- & 1.43e-02 & 4.56e-03 & 5.17e-03 & 1.53e-02 \\
$^{85}\mathrm{Sr}$ & stable & --- & 1.43e-02 & 4.56e-03 & 5.17e-03 & 1.53e-02 \\
$^{110}\mathrm{In}$ & stable & --- & 4.57e-03 & 2.57e-03 & 1.29e-02 & 5.80e-03 \\
$^{32}\mathrm{P}$ & stable & --- & 1.43e-02 & 4.56e-03 & 2.58e-03 & 1.13e-02 \\
$^{32}\mathrm{S}$ & stable & --- & 1.43e-02 & 4.56e-03 & 2.58e-03 & 1.13e-02 \\
$^{91}\mathrm{Nb}$ & stable & --- & 1.15e-02 & 4.07e-03 & 5.17e-03 & 1.53e-02 \\
$^{33}\mathrm{S}$ & stable & --- & 1.15e-02 & 4.07e-03 & 5.17e-03 & 1.53e-02 \\
$^{44}\mathrm{Ti}$ & stable & --- & 1.15e-02 & 4.07e-03 & 5.17e-03 & 1.53e-02 \\
$^{105}\mathrm{Ag}$ & stable & --- & --- & --- & 7.75e-03 & 1.92e-02 \\
$^{106}\mathrm{Ag}$ & stable & --- & --- & --- & 7.75e-03 & 1.92e-02 \\
$^{26}\mathrm{Al}$ & stable & --- & --- & --- & 7.75e-03 & 1.92e-02 \\
$^{102}\mathrm{Cd}$ & stable & --- & --- & --- & 7.75e-03 & 1.92e-02 \\
$^{62}\mathrm{Co}$ & stable & --- & --- & --- & 7.75e-03 & 1.92e-02 \\
$^{70}\mathrm{Ge}$ & stable & --- & --- & --- & 7.75e-03 & 1.92e-02 \\
$^{9}\mathrm{He}$ & stable & --- & --- & --- & 7.75e-03 & 1.92e-02 \\
$^{115}\mathrm{In}$ & stable & --- & --- & --- & 7.75e-03 & 1.92e-02 \\
$^{117}\mathrm{In}$ & stable & --- & --- & --- & 7.75e-03 & 1.92e-02 \\
$^{119}\mathrm{In}$ & stable & --- & --- & --- & 7.75e-03 & 1.92e-02 \\
$^{82}\mathrm{Kr}$ & stable & --- & --- & --- & 7.75e-03 & 1.92e-02 \\
$^{16}\mathrm{Ne}$ & stable & --- & --- & --- & 7.75e-03 & 1.92e-02 \\
$^{105}\mathrm{Pd}$ & stable & --- & --- & --- & 7.75e-03 & 1.92e-02 \\
$^{99}\mathrm{Rh}$ & stable & --- & --- & --- & 7.75e-03 & 1.92e-02 \\
$^{99}\mathrm{Ru}$ & stable & --- & --- & --- & 7.75e-03 & 1.92e-02 \\
$^{134}\mathrm{Sb}$ & stable & --- & --- & --- & 7.75e-03 & 1.92e-02 \\
$^{124}\mathrm{Sn}$ & stable & --- & --- & --- & 7.75e-03 & 1.92e-02 \\
$^{96}\mathrm{Tc}$ & stable & --- & --- & --- & 7.75e-03 & 1.92e-02 \\
$^{119}\mathrm{Xe}$ & stable & --- & --- & --- & 7.75e-03 & 1.92e-02 \\
$^{108}\mathrm{Ag}$ & stable & --- & 1.29e-02 & 4.32e-03 & 2.58e-03 & 1.13e-02 \\
$^{109}\mathrm{Ag}$ & stable & --- & 1.03e-02 & 3.85e-03 & 5.17e-03 & 1.53e-02 \\
$^{102}\mathrm{Ag}$ & stable & --- & --- & --- & 5.17e-03 & 1.53e-02 \\
$^{75}\mathrm{Br}$ & stable & --- & --- & --- & 5.17e-03 & 1.53e-02 \\
$^{78}\mathrm{Br}$ & stable & --- & --- & --- & 5.17e-03 & 1.53e-02 \\
$^{103}\mathrm{Cd}$ & stable & --- & --- & --- & 5.17e-03 & 1.53e-02 \\
$^{113}\mathrm{Cd}$ & stable & --- & --- & --- & 5.17e-03 & 1.53e-02 \\
$^{122}\mathrm{Cs}$ & stable & --- & --- & --- & 5.17e-03 & 1.53e-02 \\
$^{59}\mathrm{Fe}$ & stable & --- & --- & --- & 5.17e-03 & 1.53e-02 \\
$^{69}\mathrm{Ga}$ & stable & --- & --- & --- & 5.17e-03 & 1.53e-02 \\
$^{70}\mathrm{Ga}$ & stable & --- & --- & --- & 5.17e-03 & 1.53e-02 \\
$^{67}\mathrm{Ge}$ & stable & --- & --- & --- & 5.17e-03 & 1.53e-02 \\
$^{68}\mathrm{Ge}$ & stable & --- & --- & --- & 5.17e-03 & 1.53e-02 \\
$^{73}\mathrm{Ge}$ & stable & --- & --- & --- & 5.17e-03 & 1.53e-02 \\
$^{116}\mathrm{In}$ & stable & --- & --- & --- & 5.17e-03 & 1.53e-02 \\
$^{120}\mathrm{In}$ & stable & --- & --- & --- & 5.17e-03 & 1.53e-02 \\
$^{121}\mathrm{In}$ & stable & --- & --- & --- & 5.17e-03 & 1.53e-02 \\
$^{77}\mathrm{Kr}$ & stable & --- & --- & --- & 5.17e-03 & 1.53e-02 \\
$^{27}\mathrm{Mg}$ & stable & --- & --- & --- & 5.17e-03 & 1.53e-02 \\
$^{91}\mathrm{Mo}$ & stable & --- & --- & --- & 5.17e-03 & 1.53e-02 \\
$^{95}\mathrm{Mo}$ & stable & --- & --- & --- & 5.17e-03 & 1.53e-02 \\
$^{10}\mathrm{N}$ & stable & --- & --- & --- & 5.17e-03 & 1.53e-02 \\
$^{89}\mathrm{Nb}$ & stable & --- & --- & --- & 5.17e-03 & 1.53e-02 \\
$^{90}\mathrm{Nb}$ & stable & --- & --- & --- & 5.17e-03 & 1.53e-02 \\
$^{93}\mathrm{Nb}$ & stable & --- & --- & --- & 5.17e-03 & 1.53e-02 \\
$^{22}\mathrm{Ne}$ & stable & --- & --- & --- & 5.17e-03 & 1.53e-02 \\
$^{106}\mathrm{Pd}$ & stable & --- & --- & --- & 5.17e-03 & 1.53e-02 \\
$^{110}\mathrm{Pd}$ & stable & --- & --- & --- & 5.17e-03 & 1.53e-02 \\
$^{99}\mathrm{Pd}$ & stable & --- & --- & --- & 5.17e-03 & 1.53e-02 \\
$^{101}\mathrm{Rh}$ & stable & --- & --- & --- & 5.17e-03 & 1.53e-02 \\
$^{107}\mathrm{Rh}$ & stable & --- & --- & --- & 5.17e-03 & 1.53e-02 \\
$^{95}\mathrm{Rh}$ & stable & --- & --- & --- & 5.17e-03 & 1.53e-02 \\
$^{101}\mathrm{Ru}$ & stable & --- & --- & --- & 5.17e-03 & 1.53e-02 \\
$^{102}\mathrm{Ru}$ & stable & --- & --- & --- & 5.17e-03 & 1.53e-02 \\
$^{93}\mathrm{Ru}$ & stable & --- & --- & --- & 5.17e-03 & 1.53e-02 \\
$^{27}\mathrm{Si}$ & stable & --- & --- & --- & 5.17e-03 & 1.53e-02 \\
$^{31}\mathrm{Si}$ & stable & --- & --- & --- & 5.17e-03 & 1.53e-02 \\
$^{128}\mathrm{Sn}$ & stable & --- & --- & --- & 5.17e-03 & 1.53e-02 \\
$^{129}\mathrm{Sn}$ & stable & --- & --- & --- & 5.17e-03 & 1.53e-02 \\
$^{84}\mathrm{Sr}$ & stable & --- & --- & --- & 5.17e-03 & 1.53e-02 \\
$^{93}\mathrm{Tc}$ & stable & --- & --- & --- & 5.17e-03 & 1.53e-02 \\
$^{94}\mathrm{Tc}$ & stable & --- & --- & --- & 5.17e-03 & 1.53e-02 \\
$^{113}\mathrm{Te}$ & stable & --- & --- & --- & 5.17e-03 & 1.53e-02 \\
$^{86}\mathrm{Y}$ & stable & --- & --- & --- & 5.17e-03 & 1.53e-02 \\
$^{86}\mathrm{Zr}$ & stable & --- & --- & --- & 5.17e-03 & 1.53e-02 \\
$^{87}\mathrm{Zr}$ & stable & --- & --- & --- & 5.17e-03 & 1.53e-02 \\
$^{110}\mathrm{Ag}$ & stable & --- & --- & --- & 2.58e-03 & 1.13e-02 \\
$^{111}\mathrm{Ag}$ & stable & --- & --- & --- & 2.58e-03 & 1.13e-02 \\
$^{71}\mathrm{As}$ & stable & --- & --- & --- & 2.58e-03 & 1.13e-02 \\
$^{73}\mathrm{As}$ & stable & --- & --- & --- & 2.58e-03 & 1.13e-02 \\
$^{75}\mathrm{As}$ & stable & --- & --- & --- & 2.58e-03 & 1.13e-02 \\
$^{128}\mathrm{Ba}$ & stable & --- & --- & --- & 2.58e-03 & 1.13e-02 \\
$^{13}\mathrm{Be}$ & stable & --- & --- & --- & 2.58e-03 & 1.13e-02 \\
$^{76}\mathrm{Br}$ & stable & --- & --- & --- & 2.58e-03 & 1.13e-02 \\
$^{77}\mathrm{Br}$ & stable & --- & --- & --- & 2.58e-03 & 1.13e-02 \\
$^{79}\mathrm{Br}$ & stable & --- & --- & --- & 2.58e-03 & 1.13e-02 \\
$^{81}\mathrm{Br}$ & stable & --- & --- & --- & 2.58e-03 & 1.13e-02 \\
$^{46}\mathrm{Ca}$ & stable & --- & --- & --- & 2.58e-03 & 1.13e-02 \\
$^{115}\mathrm{Cd}$ & stable & --- & --- & --- & 2.58e-03 & 1.13e-02 \\
$^{117}\mathrm{Cd}$ & stable & --- & --- & --- & 2.58e-03 & 1.13e-02 \\
$^{42}\mathrm{Cl}$ & stable & --- & --- & --- & 2.58e-03 & 1.13e-02 \\
$^{120}\mathrm{Cs}$ & stable & --- & --- & --- & 2.58e-03 & 1.13e-02 \\
$^{123}\mathrm{Cs}$ & stable & --- & --- & --- & 2.58e-03 & 1.13e-02 \\
$^{58}\mathrm{Cu}$ & stable & --- & --- & --- & 2.58e-03 & 1.13e-02 \\
$^{67}\mathrm{Cu}$ & stable & --- & --- & --- & 2.58e-03 & 1.13e-02 \\
$^{69}\mathrm{Cu}$ & stable & --- & --- & --- & 2.58e-03 & 1.13e-02 \\
$^{60}\mathrm{Fe}$ & stable & --- & --- & --- & 2.58e-03 & 1.13e-02 \\
$^{61}\mathrm{Fe}$ & stable & --- & --- & --- & 2.58e-03 & 1.13e-02 \\
$^{63}\mathrm{Fe}$ & stable & --- & --- & --- & 2.58e-03 & 1.13e-02 \\
$^{68}\mathrm{Ga}$ & stable & --- & --- & --- & 2.58e-03 & 1.13e-02 \\
$^{116}\mathrm{I}$ & stable & --- & --- & --- & 2.58e-03 & 1.13e-02 \\
$^{117}\mathrm{I}$ & stable & --- & --- & --- & 2.58e-03 & 1.13e-02 \\
$^{118}\mathrm{I}$ & stable & --- & --- & --- & 2.58e-03 & 1.13e-02 \\
$^{105}\mathrm{In}$ & stable & --- & --- & --- & 2.58e-03 & 1.13e-02 \\
$^{106}\mathrm{In}$ & stable & --- & --- & --- & 2.58e-03 & 1.13e-02 \\
$^{122}\mathrm{In}$ & stable & --- & --- & --- & 2.58e-03 & 1.13e-02 \\
$^{123}\mathrm{In}$ & stable & --- & --- & --- & 2.58e-03 & 1.13e-02 \\
$^{125}\mathrm{In}$ & stable & --- & --- & --- & 2.58e-03 & 1.13e-02 \\
$^{43}\mathrm{K}$ & stable & --- & --- & --- & 2.58e-03 & 1.13e-02 \\
$^{75}\mathrm{Kr}$ & stable & --- & --- & --- & 2.58e-03 & 1.13e-02 \\
$^{76}\mathrm{Kr}$ & stable & --- & --- & --- & 2.58e-03 & 1.13e-02 \\
$^{81}\mathrm{Kr}$ & stable & --- & --- & --- & 2.58e-03 & 1.13e-02 \\
$^{56}\mathrm{Mn}$ & stable & --- & --- & --- & 2.58e-03 & 1.13e-02 \\
$^{58}\mathrm{Mn}$ & stable & --- & --- & --- & 2.58e-03 & 1.13e-02 \\
$^{90}\mathrm{Mo}$ & stable & --- & --- & --- & 2.58e-03 & 1.13e-02 \\
$^{93}\mathrm{Mo}$ & stable & --- & --- & --- & 2.58e-03 & 1.13e-02 \\
$^{96}\mathrm{Mo}$ & stable & --- & --- & --- & 2.58e-03 & 1.13e-02 \\
$^{23}\mathrm{Na}$ & stable & --- & --- & --- & 2.58e-03 & 1.13e-02 \\
$^{24}\mathrm{Na}$ & stable & --- & --- & --- & 2.58e-03 & 1.13e-02 \\
$^{25}\mathrm{Na}$ & stable & --- & --- & --- & 2.58e-03 & 1.13e-02 \\
$^{87}\mathrm{Nb}$ & stable & --- & --- & --- & 2.58e-03 & 1.13e-02 \\
$^{94}\mathrm{Nb}$ & stable & --- & --- & --- & 2.58e-03 & 1.13e-02 \\
$^{14}\mathrm{Ne}$ & stable & --- & --- & --- & 2.58e-03 & 1.13e-02 \\
$^{24}\mathrm{Ne}$ & stable & --- & --- & --- & 2.58e-03 & 1.13e-02 \\
$^{66}\mathrm{Ni}$ & stable & --- & --- & --- & 2.58e-03 & 1.13e-02 \\
$^{109}\mathrm{Pd}$ & stable & --- & --- & --- & 2.58e-03 & 1.13e-02 \\
$^{111}\mathrm{Pd}$ & stable & --- & --- & --- & 2.58e-03 & 1.13e-02 \\
$^{96}\mathrm{Pd}$ & stable & --- & --- & --- & 2.58e-03 & 1.13e-02 \\
$^{98}\mathrm{Pd}$ & stable & --- & --- & --- & 2.58e-03 & 1.13e-02 \\
$^{80}\mathrm{Rb}$ & stable & --- & --- & --- & 2.58e-03 & 1.13e-02 \\
$^{81}\mathrm{Rb}$ & stable & --- & --- & --- & 2.58e-03 & 1.13e-02 \\
$^{83}\mathrm{Rb}$ & stable & --- & --- & --- & 2.58e-03 & 1.13e-02 \\
$^{102}\mathrm{Rh}$ & stable & --- & --- & --- & 2.58e-03 & 1.13e-02 \\
$^{106}\mathrm{Rh}$ & stable & --- & --- & --- & 2.58e-03 & 1.13e-02 \\
$^{96}\mathrm{Rh}$ & stable & --- & --- & --- & 2.58e-03 & 1.13e-02 \\
$^{98}\mathrm{Rh}$ & stable & --- & --- & --- & 2.58e-03 & 1.13e-02 \\
$^{105}\mathrm{Ru}$ & stable & --- & --- & --- & 2.58e-03 & 1.13e-02 \\
$^{95}\mathrm{Ru}$ & stable & --- & --- & --- & 2.58e-03 & 1.13e-02 \\
$^{47}\mathrm{Sc}$ & stable & --- & --- & --- & 2.58e-03 & 1.13e-02 \\
$^{74}\mathrm{Se}$ & stable & --- & --- & --- & 2.58e-03 & 1.13e-02 \\
$^{107}\mathrm{Sn}$ & stable & --- & --- & --- & 2.58e-03 & 1.13e-02 \\
$^{127}\mathrm{Sn}$ & stable & --- & --- & --- & 2.58e-03 & 1.13e-02 \\
$^{130}\mathrm{Sn}$ & stable & --- & --- & --- & 2.58e-03 & 1.13e-02 \\
$^{77}\mathrm{Sr}$ & stable & --- & --- & --- & 2.58e-03 & 1.13e-02 \\
$^{80}\mathrm{Sr}$ & stable & --- & --- & --- & 2.58e-03 & 1.13e-02 \\
$^{81}\mathrm{Sr}$ & stable & --- & --- & --- & 2.58e-03 & 1.13e-02 \\
$^{87}\mathrm{Sr}$ & stable & --- & --- & --- & 2.58e-03 & 1.13e-02 \\
$^{100}\mathrm{Tc}$ & stable & --- & --- & --- & 2.58e-03 & 1.13e-02 \\
$^{95}\mathrm{Tc}$ & stable & --- & --- & --- & 2.58e-03 & 1.13e-02 \\
$^{97}\mathrm{Tc}$ & stable & --- & --- & --- & 2.58e-03 & 1.13e-02 \\
$^{98}\mathrm{Tc}$ & stable & --- & --- & --- & 2.58e-03 & 1.13e-02 \\
$^{116}\mathrm{Xe}$ & stable & --- & --- & --- & 2.58e-03 & 1.13e-02 \\
$^{83}\mathrm{Y}$ & stable & --- & --- & --- & 2.58e-03 & 1.13e-02 \\
$^{85}\mathrm{Y}$ & stable & --- & --- & --- & 2.58e-03 & 1.13e-02 \\
$^{87}\mathrm{Y}$ & stable & --- & --- & --- & 2.58e-03 & 1.13e-02 \\
$^{89}\mathrm{Y}$ & stable & --- & --- & --- & 2.58e-03 & 1.13e-02 \\
$^{90}\mathrm{Y}$ & stable & --- & --- & --- & 2.58e-03 & 1.13e-02 \\
$^{60}\mathrm{Zn}$ & stable & --- & --- & --- & 2.58e-03 & 1.13e-02 \\
$^{61}\mathrm{Zn}$ & stable & --- & --- & --- & 2.58e-03 & 1.13e-02 \\
$^{65}\mathrm{Zn}$ & stable & --- & --- & --- & 2.58e-03 & 1.13e-02 \\
$^{69}\mathrm{Zn}$ & stable & --- & --- & --- & 2.58e-03 & 1.13e-02 \\
$^{93}\mathrm{Zr}$ & stable & --- & --- & --- & 2.58e-03 & 1.13e-02 \\
$^{96}\mathrm{Zr}$ & stable & --- & --- & --- & 2.58e-03 & 1.13e-02 \\
$^{132}\mathrm{Ba}$ & stable & --- & 9.91e-04 & 1.19e-03 & 2.58e-03 & 1.13e-02 \\
$^{108}\mathrm{Pd}$ & stable & --- & 3.67e-04 & 7.25e-04 & 2.58e-03 & 1.13e-02 \\
$^{91}\mathrm{Zr}$ & stable & --- & 1.17e-05 & 1.29e-04 & 2.58e-03 & 1.13e-02 \\
$^{36}\mathrm{Ar}$ & stable & --- & 7.43e-08 & 1.03e-05 & 2.58e-03 & 1.13e-02 \\
$^{40}\mathrm{Ca}$ & stable & --- & 1.85e-11 & 1.63e-07 & 2.58e-03 & 1.13e-02 \\
$^{130}\mathrm{Ba}$ & stable & --- & 7.56e-04 & 1.04e-03 & --- & --- \\
$^{114}\mathrm{Cd}$ & stable & --- & 7.16e-05 & 3.20e-04 & --- & --- \\
$^{135}\mathrm{Ba}$ & stable & --- & 1.01e-05 & 1.20e-04 & --- & --- \\
$^{36}\mathrm{S}$ & stable & --- & 1.44e-09 & 1.44e-06 & --- & --- \\
$^{40}\mathrm{Ar}$ & stable & --- & 2.22e-12 & 5.63e-08 & --- & --- \\
\end{longtable}

\end{document}